\documentclass[prc,twocolumn,superscriptaddress,showpacs,twoside,floatfix]
{revtex4-1}

\usepackage{amssymb,epsfig}
\usepackage{color}

\begin{document}


\title{From Coulomb excitation cross sections to non-resonant astrophysical rates in three-body systems: $^{17}$Ne case}

\author{Yu.L.~Parfenova}
\email{parfenova@hotmail.com}
\affiliation{Flerov Laboratory of Nuclear Reactions, JINR, Dubna, RU-141980, Russia}

\author{L.V.~Grigorenko}
\affiliation{Flerov Laboratory of Nuclear Reactions, JINR, Dubna, RU-141980, Russia}
\affiliation{National Research Nuclear University ``MEPhI'',
115409 Moscow, Russia}
\affiliation{National Research Centre ``Kurchatov Institute'', Kurchatov
sq.\ 1, 123182 Moscow, Russia}

\author{I.A.~Egorova}
\affiliation{Bogoliubov Laboratory of Theoretical Physics, JINR, Dubna, RU-141980 Russia}
\affiliation{Department of Physics, Western Michigan University, Kalamazoo, MI 49008, USA}
\affiliation{SSC RF ITEP of NRC Kurchatov Institute, Institute for Theoretical and Experimental Physics, Moscow 117218, Russia}

\author{N.B.~Shulgina}
\affiliation{Bogoliubov Laboratory of Theoretical Physics, JINR, Dubna, RU-141980 Russia}
\affiliation{National Research Centre ``Kurchatov Institute'', Kurchatov
sq.\ 1, 123182 Moscow, Russia}

\author{J.S.~Vaagen}
\affiliation{Institute of Physics and Technology, University of Bergen, N-5007 Bergen, Norway}
\author{M.V.~Zhukov}
\affiliation{Department of Physics, Chalmers University of Technology, S-41296
G\"{o}teborg, Sweden}

\date{\today}

\begin{abstract}
Coulomb and nuclear dissociation of $^{17}$Ne on light and heavy targets are studied theoretically. The dipole E1 strength function is determined in a broad energy range including energies of astrophysical interest. Dependence of the strength function on different parameters of the $^{17}$Ne ground state structure and continuum dynamics is analyzed in a three-body model. The discovered dependence plays an important role for studies of the strength functions for the three-body E1 dissociation and radiative capture. The constraints on the $[s^2]/[d^2]$ configuration mixing in $^{17}$Ne and on $p$-wave interaction in the $^{15}$O+$p$ channel are imposed based on  experimental data for $^{17}$Ne Coulomb dissociation on heavy target.
\end{abstract}

\maketitle


\section{Introduction}
\label{sec:intro}


An important application of nuclear studies is the determination of the astrophysical reaction rates, which are basis for nucleosynthesis calculations. The radiative capture rates has two qualitatively different contributions: resonant and non-resonant. For studies of the \emph{resonant} radiative capture rates only the basic information about resonances is required: resonant energy, particle and gamma widths (for simplicity we discuss below the situation where one particle and one gamma channel dominate):
\begin{equation}
\langle \sigma_{\mathrm{part},\gamma} v \rangle (T) \sim \frac{1}{T^{3n/2}}\,
\exp \left( -\frac{E_r}{kT} \right)\, \frac{ \Gamma_{\gamma} \Gamma_{\mathrm{part}} } {\Gamma_{\mathrm{tot}}}\,,
\label{eq:res-rate-s}
\end{equation}
where $E_r$ is the resonance position, $\Gamma_{\gamma}$ and $\Gamma_{\mathrm{part}}$ are partial widths of the resonance, decaying into gamma and particle channels \cite{Fowler:1967}, and $T$ is the temperature. The ``particle'' may here be proton, alpha, two protons, etc, and $n$ is the number of captured particles: $n=1$ for $p$, $\alpha$ and $n=2$ for $2p$ captures. It is easy to see that, with $\Gamma_{\mathrm{tot}}=\Gamma_{\gamma}+\Gamma_{\mathrm{part}}$, the astrophysical resonant rate depends only on $\Gamma_{\gamma}$ for $\Gamma_{\mathrm{part}} \gg \Gamma_{\gamma}$ or only on $\Gamma_{\mathrm{part}}$ for $\Gamma_{\mathrm{part}} \ll \Gamma_{\gamma}$. Measurements of values needed for resonance rate determination could be complicated but it is a realistic task in most systems of interest. The situation is totally different for \emph{nonresonant} radiative capture rates. The cross sections of the reciprocal reactions of photo and Coulomb dissociation can be used for the nonresonant rate determination. However, the direct measurements of such cross sections could be not feasible for the low energies of astrophysical interest. The direct cross section measurements are also not feasible for three-body capture since such processes (practically simultaneous collisions of three particles) become noticeable only at stellar densities and energies.

\begin{figure}
\begin{center}
\includegraphics[width=0.47\textwidth]{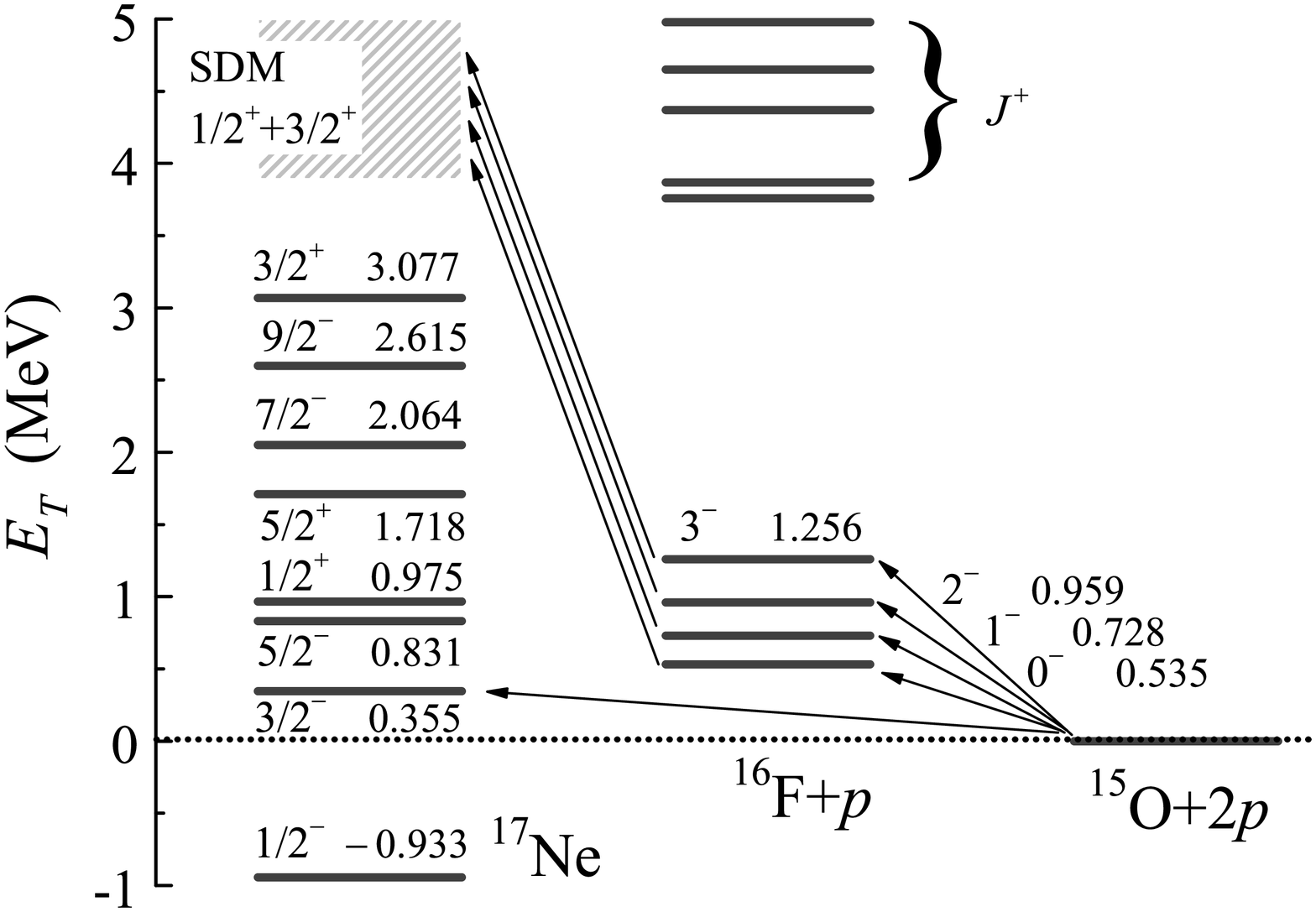}
\includegraphics[width=0.47\textwidth]{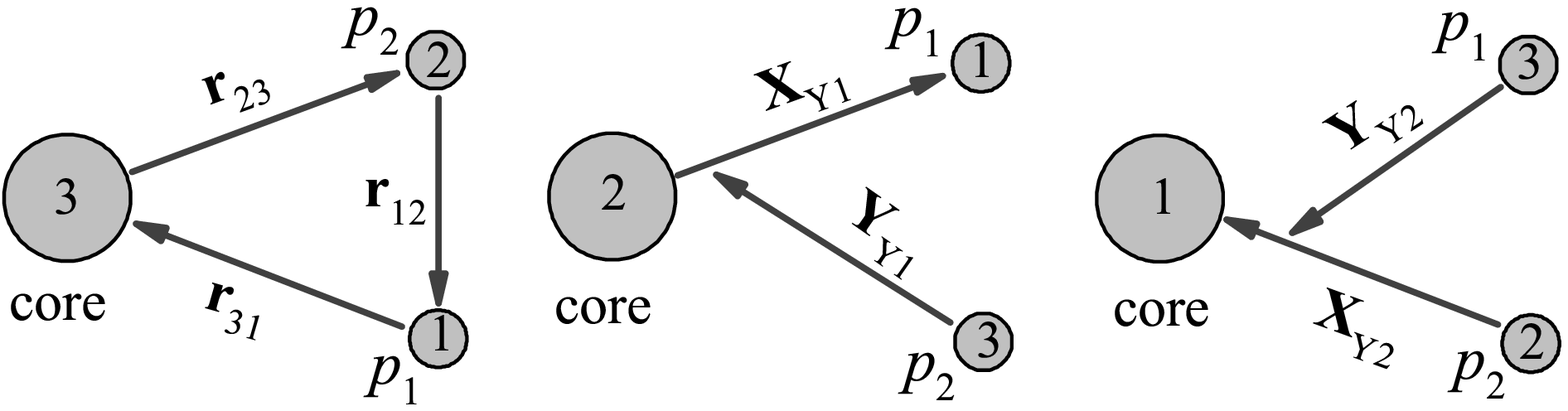}
\end{center}
\caption{The level schemes of $^{17}$Ne, $^{16}$F and the coordinate systems for three-body representation of $^{17}$Ne used in this work. Arrows in the upper panel illustrate the processes of astrophysical relevance: (i) direct resonant $2p$ capture via the first excited $3/2^-$ state of $^{17}$Ne and (ii) direct two-step non-resonant capture via soft dipole mechanism  (SDM, $1/2^+$ and $3/2^+$ quantum numbers) with strength functions peaked above 3.5 MeV.}
\label{fig:schemes}
\end{figure}

The astrophysical problem of $^{17}$Ne has two major aspects. The \emph{resonant} radiative capture rate for $^{15}$O+$p$+$p \rightarrow ^{17}$Ne+$\gamma$ at the temperatures of astrophysical relevance, crucially depends on the $2p$ width of the $3/2^-$ first excited state of $^{17}$Ne, see Ref.\ \cite{Grigorenko:2005a} and Fig.\ \ref{fig:schemes}. The direct experimental observation of the $2p$-decay of the $3/2^-$ state was attempted several times in the papers \cite{Chromik:2002,Sharov:2017} providing improving limits $\Gamma_{2p}/\Gamma_{\gamma} \leq 7.7 \times 10^{-3}$ and  $\Gamma_{2p}/\Gamma_{\gamma} \leq 1.6 \times 10^{-4}$, respectively. The theoretical calculations \cite{Grigorenko:2007} predict $\Gamma_{2p}/\Gamma_{\gamma} \sim (0.9-2.5) \times 10^{-6}$. If the theoretically predicted value is realistic then significant improvement of experiment is required to make the direct measurements of such a  value possible. Recently an opinion was expressed in Ref.\ \cite{Casal:2016} that the resonance rate is not important because it is negligibly small compared to the non-resonant contribution to the rate. We demonstrate in this work that the results of \cite{Casal:2016} for non-resonant rate are incorrect, and that the issue of a balance between resonant and non-resonant contributions to the rate at different temperatures pointed out by us in Ref.\ \cite{Grigorenko:2006} remains important.

The \emph{nonresonant} radiative capture rate strongly depends on the distribution of the non-resonant E1 strength in the spectrum of $^{17}$Ne \cite{Grigorenko:2006},
\begin{eqnarray}
\left\langle \sigma _{2p,\gamma }v\right\rangle &= & \left( \frac{A_1+A_2+A_3}{A_1 A_2 A_3}\right)
^{3/2}\left( \frac{2\pi }{mkT}\right) ^{3}\;\frac{2J_f+1}{2(2J_i+1)} \nonumber
\\
& \times & \int dE \, \frac{16\pi}{9}\, E_{\gamma }^{3}\;
\frac{dB_{E1}(E)}{dE} \exp \left[ -\frac{E}{kT}\right] \,,
\label{eq:nonres-rate}
\end{eqnarray}
where $J_i$ and $J_f$  are spins of the $^{15}$O and $^{17}$Ne g.s.,\
respectively. Note that $dB_{E1}/dE$ in Eq.\ (\ref{eq:nonres-rate}) is the E1 strength function for the reciprocal process of $^{17}$Ne dissociation. The above expression somewhat differs from that in \cite{Grigorenko:2006} (the factor $e^2$ is moved to strength function definition).

The shape of the E1 strength function in $^{17}$Ne is governed by the so-called soft dipole mode (SDM) existing in this nucleus. Properties of the SDM in the three-body systems were investigated in details in the Borromean $2n$ halo nuclei $^{6}$He and $^{11}$Li \cite[and Refs.\ therein]{Aumann:2005,Aumann:2013}. The sensitivity of the astrophysical non-resonant radiative capture rate to the SDM in $^{17}$Ne was discussed by some of us in Ref.\ \cite{Grigorenko:2006}. In the present paper we further elaborate on this problem using the recently available data on Coulomb dissociation of $^{17}$Ne \cite{Marganiec:2016,Wamers:2018} on heavy (Pb) target. Our aim is to clarify which type of information should be necessary and sufficient for determination of the three-body non-resonant astrophysical rate from the experimental data.

In a generally accepted approach (e.g.\ \cite{Aumann:2005}) for derivation of three-body radiative capture rate, the E1 strength function from the Coulomb dissociation cross section is used. The main problem here is that the experimental determination of the strength function is feasible at energies above several hundreds keV. For astrophysics the energies under several tens of keV are typically important. Thus, we need a reliable procedure to ``extrapolate'' correctly, to extreme low energies, the properties of the strength function observed experimentally in the range from about one to several MeV.

The unit system $\hbar=c=1$ is used in this work.

\section{Theoretical model}
\label{sec:model}


For calculations of energy spectrum and correlations in the three-body dissociation of $^{17}$Ne projectiles on lead, silicon and carbon targets at energy 500 AMeV, several approaches are combined. The three-body hyperspherical harmonic (HH) method is used for the $^{17}$Ne ground state (g.s) calculations \cite{Grigorenko:2003}. A Green's function approach with simplified three-body Hamiltonian is applied to calculations of $^{17}$Ne continuum, populated by E1 transitions. The Bertulani-Baur model \cite{Bertulani:1988b} along with the Glauber model \cite{Bertsch:1998} are used for description of Coulomb and nuclear dissociation.


\subsection{Three-body bound state}


The bound $^{17}$Ne g.s.\ wave function (WF) $\Psi_{\text{g.s.}}$ is obtained in a $^{15}$O+$p$+$p$ model by solving the three-body Schr\"odinger equation
\begin{eqnarray}
(\hat{H}_3 - E_b)\Psi_{\text{gs}} = 0 \,, \nonumber \\
\hat{H}_3 = \hat{T}_3 + V_{12}(\mathbf{r}_{12}) + V_{23}(\mathbf{r}_{23}) + V_{31}(\mathbf{r}_{31}) + V_3(\rho) \,,
\label{eq:schred-bs}
\end{eqnarray}
This equation is solved by using the HH method \cite{Grigorenko:2003}.
The $^{17}$Ne g.s.\ WF used in this work has previously been obtained in Ref.\ \cite{Grigorenko:2003} and further tested in the works \cite{Grigorenko:2005,Grigorenko:2007a} against different observables. The major features of this WF comprise binding energy $E_b = - 0.933$ MeV and nuclear structure with strong $[s^2]/[d^2]$ configuration mixing ($\sim 50\%$ of $[s^2]$ configuration). This structure is defined by the two-body resonant states in the $^{15}$O+$p$ channel: $s$-wave $0^-$ and $1^-$, $d$-wave $2^-$ and $3^-$, at 0.535, 0.728, 0.959, and 1.256 MeV, respectively, see Fig.\ \ref{fig:schemes}. Attractive interaction in the $p$-wave was assumed in our calculations \cite{Grigorenko:2003,Grigorenko:2005,Grigorenko:2007a}. However, this interaction is relatively weak so that there is no single-particle $p$-wave state in $^{16}$F below 3 MeV (which would contradict experimental data on the $^{16}$F and $^{16}$N spectra).


\subsection{Soft E1 strength function with simplified three-body Green's function}


The continuum WFs of the positive parity states in $^{17}$Ne, populated in E1 transition, are obtained by means of a  Green's function method \cite{Grigorenko:2006},
\begin{equation}
\Psi_{3E_T,M'm}^{JM(+)} = \hat G_{3E_T}^{(+)} \, {\cal O}_{E1,m} \, \Psi_{\text{gs}}^{J'M'} \,.
\end{equation}
The continuum in $^{17}$Ne is populated in the soft E1 excitation (SDM) described by the dipole operators in the cluster form:
\[
{\cal O}_{\text{E1},m} = \sqrt{\frac{3}{4\pi}} \sum \nolimits_i e \, Z_i r_i \,  Y_{1m}(\hat {\bf r}_i) \,,
\]
where $\textbf{r}_i$ and $Z_i$ are coordinates and charge numbers of the individual clusters. The Green's function $G_{3E_T}^{(+)}$ corresponds to a simplified three-body Hamiltonian
\begin{equation}
\hat{H}'_3 = \hat{T}_3 + V_{12}(\mathbf{X}_{Y_2}) + V_{Y}(\mathbf{Y}_{Y_2})\,.
\label{eq:ham-sim}
\end{equation}
The Green's function for Hamiltonians containing terms depending on separated Jacobi variables is available in compact analytical form
\begin{equation}
\hat G_{3E_T}^{(+)}({\bf XY,X'Y'}) =  \int \limits_0^{E_T}
\frac{d E_x}{2 \pi i} \, \hat G_{E_x}^{(+)}({\bf X,X'}) \, \hat G_{E_y}^{(+)}({\bf Y,Y'})\,,
\end{equation}
where $E_x$ and $E_y$, $E_x+E_y=E_T$ are energies in the ``X'' and ``Y'' Jacobi subsystems, see Fig.\ \ref{fig:schemes}. The operators $\hat G_{E}^{(+)}$ are  ordinary two-body Green's functions for the corresponding channels. This method takes into account exactly the final-state interaction for one of three pairs of clusters only. This is a reasonable approximation since for ``non-natural'' parity states of the core+$N$+$N$ system only one of pairwise interactions (core+$N$ with ``natural'' parity) is essential for description of the decay dynamics, see Refs.\ \cite{Grigorenko:2006,Fomichev:2012} for details.

As far as we are interested in population of continuum states with definite $J^{\pi}$, and the spectrum in the ``X'' subsystem contains a number of $^{16}$F states also with definite $j_x^{\pi_x}$, the actual form of the Green's function should take into account the angular momentum coupling as follows:
\begin{widetext}
\begin{eqnarray}
\hat G_{3E_T}^{JMM'(+)}({\bf XY,X'Y'}) = \sum_{l_x S_x j_x l_y j_y} \int \limits_0^{E_T}
\frac{d E_x}{2 \pi i} \, \frac{2M_x}{k_x X X'}
\left \{
\begin{array}{l}
 f_{l_x S_x j_x}(k_xX) h^{(+)}_{l_x S_x j_x}(k_xX'),X < X'  \\
 h^{(+)}_{l_x S_x j_x}(k_xX) f_{l_x S_x j_x}(k_xX') ,X > X'
\end{array}
\right \}
\frac{2M_y}{k_y Y Y'} \qquad \\
\times \left \{
\begin{array}{l}
 f_{l_y j_y J}(k_yY) h^{(+)}_{l_y j_y J}(k_yY'),Y < Y'  \\
 h^{(+)}_{l_y j_y J}(k_yY) f_{l_y j_y J}(k_yY') ,Y > Y'
\end{array}
\right \}
[Y_{l_y} \otimes [[Y_{l_x}\otimes [S_1 \otimes S_2]_{S_x}]_{j_x} \otimes S_3]_{j_y}]_{JM} \, [Y'_{l_y} \otimes [[Y'_{l_x}\otimes [S'_1 \otimes S'_2]_{S_x}]_{j_x} \otimes S'_3]_{j_y}]_{JM'} \,.\nonumber
\end{eqnarray}
\end{widetext}
Here we employ a kind of $ls$ coupling scheme, where the $^{16}$F states are characterized by quantum numbers $\{l_x,S_x,j_x\}$, where $S_x$ is total spin of $^{15}$O and one of the protons. Such scheme was employed in the original work Ref.\ \cite{Grigorenko:2003}. For consistency an analogous $ls$ coupling scheme is used in the ``Y'' subsystem with spin $j_y$ formed by total $^{16}$F spin and spin of the second proton.

The functions $f(r)$ and $h^{(+)}(r)$ are eigenfunctions of
sub-Hamiltonians of the Jacobi subsystems
\begin{eqnarray}
\hat{H}_{x}-E_x & = &  \hat{T}_{x}+V_{x}(X)-E_x \,, \nonumber \\
\hat{H}_{y}-E_y &  = & \hat{T}_{y}+V_{y}(Y)-(E_T-E_x)\,,  \nonumber
\end{eqnarray}
normalized by the asymptotic conditions at large radius $r$ as
\begin{eqnarray}
f_{l S j}(kr)  & \rightarrow & e^{i\delta_{lSj}}\;\left[  G_{l}(kr)\sin(\delta
_{lSj})+F_{l}(kr)\cos(\delta_{lSj})\right]\,, \nonumber \\
h_{lSj}^{(+)}(kr)  & \rightarrow & G_{l}(kr)+iF_{l}(kr)\,, \nonumber
\end{eqnarray}
where $F_l$ and $G_l$ are the regular and irregular (at the origin) Coulomb radial WFs.

The three-body Green's function in the above form has definite total spin-parity and contains states with  definite spin-parity in the $^{16}$F subsystem. However, it does not have definite symmetry for permutation of protons. This can be cured by explicit symmetrization of the Green's functions constructed in different Jacobi systems, see Fig.\ \ref{fig:schemes}:
\begin{eqnarray}
\hat G_{3E_T}^{JMM'(+)} & =  & \hat G_{3E_T}^{JMM'(+)}({\bf X}_{Y_1}{\bf Y}_{Y_1},{\bf X}'_{Y_1} {\bf Y}'_{Y_1}) \nonumber \\
& +  & \hat G_{3E_T}^{JMM'(+)}({\bf X}_{Y_2}{\bf Y}_{Y_2},{\bf X}'_{Y_2} {\bf Y}'_{Y_2}) \, .
\end{eqnarray}

The strength function $d B_{\text{E1}}/d E$  of the E1 Coulomb excitation is expressed via the flux $j$ induced by the $^{17}$Ne continuum wave function $\Psi_3^{(+)}$ through a remote surface $S$ as
\begin{eqnarray}
\frac{d B_{E1}}{d E_T} = \frac{1}{2\pi}  \sum \nolimits _J j_J \,, \nonumber \\
j_J = \frac{1}{2J'+1}\sum \nolimits _{MM'm} \left. \left \langle \Psi_{3E_T,M'm}^{JM(+)}| \hat{j} | \Psi_{3E_T,M'm}^{JM(+)} \right \rangle \right|_{S}\,.
\label{eq:dbde}
\end{eqnarray}

\begin{figure}
\begin{center}
\includegraphics[width=0.47\textwidth]{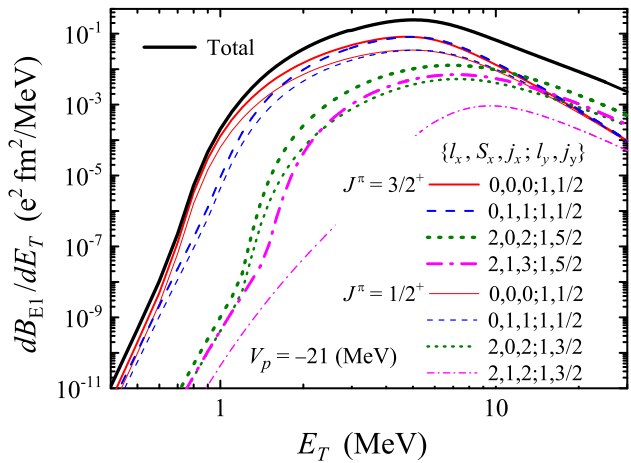}
\end{center}
\caption{Decompositions of the E1 strength function over partial contributions. The calculations are done with $V_p= -21$ MeV and the $^{17}$Ne g.s.\ WF with $\sim 50\%$ of the $[s^2]$ configuration.}
\label{fig:sfe1-v21-all}
\end{figure}

An example of the E1 strength function calculation is provided in Fig.\ \ref{fig:sfe1-v21-all}. The strength function decomposition indicates the contributions of the final states with $J^{\pi}$ equal $1/2^+$ and $3/2^+$ as well as the partial contributions of the major $\{l_x,S_x,j_x,l_y,j_y \}$ configurations. There are kinks in the partial contributions of different components between 0.5 and 1.5 MeV. Their energies correlate with energies of relevant $^{16}$F resonance states and thus signifies transitions from a true three-body dissociation regime to the dissociation proceeding ``semi-sequentially'' via different two-body resonant states in $^{16}$F.

In the Bertulani-Baur model \cite{Bertulani:1988b} the Coulomb excitation cross section is expressed via the electromagnetic strength function. The cross section $\sigma_{E_{\lambda}}$ with multipolarity $E_{\lambda}$ is
\begin{eqnarray}
\frac{d^3 \sigma_{E_{\lambda}}}{d E_T d b } =
 \, \frac{(2\pi)^3( \lambda +1)}{ \lambda [(2\lambda+1)!!]^2}
\left( E_{T}+E_{b} \right)^{2\lambda-1} \frac{d B_{E \lambda}} {d E_T} \nonumber \\
\times \, \frac{d n_{E_\lambda}}{d b} \, F_{\text{abs}} (b) \,,  \quad
\label{eq:bert-baur}
\end{eqnarray}
where $E_T$ is the energy above the $2p$-emission threshold, $E_b=0.933$ MeV is the three-body binding energy of the $^{17}$Ne $1/2^-$ ground state, and $b$ is the impact parameter of the whole three-body system. The function $d n_{E_\lambda}/d b$ is the virtual photon spectrum defined analytically \cite{Bertulani:1988b}
\begin{eqnarray}
\frac{d n_{\pi \lambda}}{d b} = 2 \pi b \, e^2 Z^2_{\text{targ}} \, \left ( \frac{\omega}{\gamma v} \right)^2 \frac{\lambda [(2\lambda+1)!!]^2}{(2\pi)^3( \lambda +1) } \qquad \nonumber \\
\times \sum_m |G_{\pi \lambda m} (v)|^2 \, K^2_m \left ( \frac{\omega b}{\gamma v} \right)\, , \qquad \\
G_{E11}=-G_{E1-1}= \frac{\sqrt{8 \pi}}{3 v} \, , \quad G_{E10}= - i \frac{4 \sqrt{ \pi}}{3 \gamma v} \, . \qquad \nonumber
\label{eq:virt-ph-sp}
\end{eqnarray}
Thus, the Coulex cross section is separated into the part depending on reaction mechanism and the part depending on structure and continuum properties of the system of interest.

The factor $F_{\text{abs}}(b)$ in Eq.\ (\ref{eq:bert-baur}) takes into account the nuclear absorption. In the  Bertulani-Baur model it is conventionally approximated by the stepwise function $\theta(b-b_{\min})$ at a minimal impact parameter (corresponding to the grazing angle). This minimal impact parameter for the lead target was taken as $b_{\min} = 9.7$ fm in Ref.\ \cite{Grigorenko:2006}, see also Fig.\ \ref{fig:cross-sect-b}. In this work we perform deeper studies of the nuclear interactions to make this aspect of calculations more precise and also to clarify the question of possible importance of the Coulomb/nuclear interference for this process. This is discussed in the next Section.


\subsection{Nuclear interaction model}

\begin{figure}
\begin{center}
\includegraphics[width=0.47\textwidth]{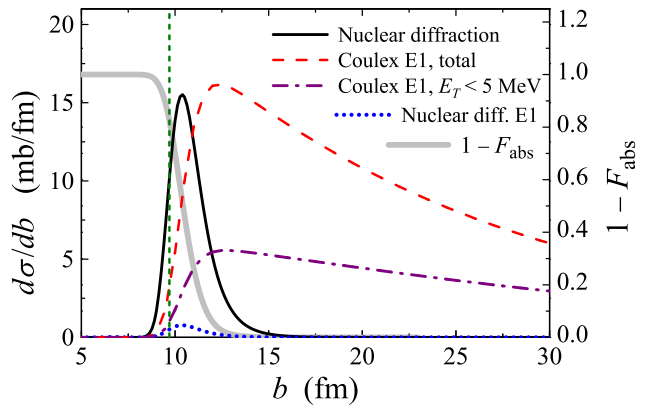}
\end{center}
\caption{The eikonal model cross sections of the nuclear, E1 nuclear, and E1 Coulomb dissociation on the lead target for 500 AMeV $^{17}$Ne beam as function of the impact parameter $b$ are shown opposite left axis by the solid, dotted and dashed curves, respectively. The contribution of the $^{17}$Ne low excitation energy region to the E1 Coulomb cross sections is illustrated by the dash-dotted curve. Calculations are done with $V_p= -21$ MeV and $^{17}$Ne g.s.\ WF with $\sim 50\%$ of the $[s^2]$ configuration. The absorption function $F_{\text{abs}}(b)$ is shown opposite right axis by the thick gray line.}
\label{fig:cross-sect-b}
\end{figure}

In this work we employed  a smooth absorption function $F_{\text{abs}}(b)$, see Fig.\ \ref{fig:cross-sect-b}. It is defined in the eikonal approximation of the Glauber model \cite{Bertsch:1998} as:
\begin{eqnarray}
F_{\text{abs}}(b)= \langle \Psi_{\text{gs}} |~|
\textstyle \prod_i S_i|^2 ~|\Psi_{\text{gs}} \rangle \, , \nonumber \\
\textstyle \prod_i S_i = S_1(\mathbf{b},\mathbf{r}_1) \, S_2(\mathbf{b},\mathbf{r}_2) \, S_3(\mathbf{b},\mathbf{r}_3) \,, \nonumber
\end{eqnarray}
where $S_i$ are the eikonal S-matrices (profile functions) of individual projectile clusters $i=\{ p, p,^{15}\text{O} \}$ with cm coordinates $\mathbf{r}_i$. The profile functions are expressed via the trajectory integrals of the effective (projectile) cluster-target interactions,
\[
S_{i}({\bf b},{\bf r}_i)=\exp \left[ -\frac{i}{v}\int\limits_{-\infty
}^{\infty }dz \, V_{i t}\left( \sqrt{({\bf b} - {\bf r}_i)^{2}+z^{2}}
\right) \right] \, .
\]
All the information on the cluster-target interaction is contained in the interaction potential $V_{i t}$. For the $^{15}$O core it was obtained as double folding of effective $NN$ interaction potentials Ref.\ \cite{DeVries:1987} with cluster and target densities, see also Refs.\ \cite{Hencken:1996,Bertsch:1998} for the details. The potentials $V_{i t}(r)$ for the valence protons are generated from the free nucleon-nucleon interaction potential \cite{Ray:1979,Charagi:1990}. The set of parameters for $^{17}$Ne calculations was obtained in the papers \cite{Parfenova:2000,Grigorenko:2005,Parfenova:2006}. It allows to reproduce the total interaction cross sections for $^{17}$Ne, $^{15}$O and cross sections of proton and two-proton removal from $^{17}$Ne in light targets.

Fig.\ \ref{fig:cross-sect-b} shows that the calculated absorption function provides somewhat larger effective cut-off radius for the Coulomb dissociation cross section than the standard one corresponding to the grazing radius approximation. This leads to reduction of the calculated cross section by about $5 \%$.

In the above approach we also get opportunity to evaluate the contribution of the nuclear dissociation processes and also possible role of the Coulomb/nuclear interference. The formalism of Glauber model eikonal approximation for inelastic, diffraction, and stripping cross sections for the halo nuclei has been presented in Ref.\ \cite{Bertsch:1998}. Within this formalism, the nuclear diffraction dissociation cross sections is written as
\begin{equation}
\frac{d \sigma_{\text{dif}}}{d^2b} =
\left \langle \Psi_{\text{gs}}
| \, 1- \textstyle \prod_i S_i|^{2} | \Psi_{\text{gs}} \right \rangle
-  \bigl|  1-\left \langle
\Psi_{\text{gs}}|\textstyle  \prod_i S_i | \Psi_{\text{gs}} \right \rangle \bigr|^{2}  \, .
\label{eq:sigdif}
\end{equation}
Fig.\ \ref{fig:cross-sect-b} shows that the contribution of the above cross section is localized in the surface region of the nucleus and in this region it overlaps considerably with the Coulomb dissociation cross section.

Expression (\ref{eq:sigdif}) for the nuclear diffraction cross section includes the contribution of all excited states in the continuum. To understand possible importance of the nuclear/Coulomb interference we need to extract the E1 contribution to this cross section. This is done by including the projection operator in the calculations of $\sigma_{\text{dif}}$:
\begin{equation}
\frac{d \sigma_{\text{dif},E1}}{d^2b} =
 \left \langle \Psi_{\text{gs}}
| \, 1- \textstyle \prod_i S_i | E1 \right \rangle  \, \left \langle E1 |1-\textstyle  \prod_i S_i | \Psi_{\text{gs}} \right \rangle  \, .
\label{eq:sigdif-e1}
\end{equation}
The ``E1 projection operator'' is named by analogy with electromagnetic transitions. It is formed by the spherical functions of Jacobi vectors
\begin{equation}
| E1 \rangle = \sum \nolimits _{l_xl_y} [Y_{l_x}(\hat{X}) \otimes  Y_{l_y}(\hat{Y})]_{LM} \,,
\label{eq:project}
\end{equation}
coupled to total angular momentum $L=1$ and negative parity $(-1)^{l_x+l_y}=-1$. Components with angular momenta up to $l_x=7$ and $l_y=7$ were considered in the calculations. Examples of the Coulomb and nuclear dissociation cross sections are also provided in Table \ref{tab:crsec}. Fig.\ \ref{fig:cross-sect-b} and Table \ref{tab:crsec} show that the contribution of the ``E1 nuclear transition'' is only a small fraction of the total nuclear contribution. Thus it is evident that effects of the nuclear/Coulomb interference can be reliably neglected.

\begin{table}[b]
\caption{Cross sections (in mb) of nuclear and Coulomb excitation for relativistic $^{17}$Ne at 500 AMeV on lead, silicon, and carbon targets. The calculations are done with $V_p= -21$ MeV and $^{17}$Ne g.s.\ WF with $\sim 50\%$ of the $[s^2]$ configuration.}
\begin{ruledtabular}
\begin{tabular}[c]{lccc}
$J^{\pi}$ $(E^*$ MeV)     &  Pb    &  Si    & C       \\
\hline
$3/2^-$ (1.288)              & 9.3    &  0.74  &  0.050  \\
$5/2^-$ (1.764)              & 17.6   &  1.39  &  0.094  \\
$5/2^+$ (2.651)              & 1.56   &  0.20  &  0.029  \\
Coulomb total                &  350   &  16.1  &  3.1    \\
Coulomb soft E1              &  322   &  13.8  &  2.9    \\
Coulomb soft E1, $E_T<7$ MeV &  243   &  9.6   &  2.0    \\
Coulomb soft E1, $E_T<5$ MeV &  148   &  5.9   &  1.2    \\
Nuclear  E1                  & 1.2    & 0.5    &  0.4    \\
Nuclear total                & 35     & 13     &  12     \\
\end{tabular}
\end{ruledtabular}
\label{tab:crsec}
\end{table}


\section{Qualitative properties of the E1 strength function}
\label{sec:calc}


The two-body non-resonant E1 radiative capture process for weakly bound nuclei may be fully defined just by two parameters: (i) binding energy and (ii) asymptotic normalization coefficient (ANC). We can mention here the thoroughly investigated case of $^{7}$Be+$p \rightarrow ^8$B+$\gamma$  radiative capture \cite{Kavanagh:1969,Filippone:1983,Hass:1999,Hammache:1998,Hammache:2001}, which was especially carefully elaborated because of its connection to the Solar boron neutrino problem \cite{Mukhamedzhanov:1990,Timofeyuk:1997}.

Three-body non-resonant E1 radiative capture is a much more complicated process. Specifically for $^{17}$Ne there is strong dependence of the E1 strength function on four major aspects of nuclear structure which we demonstrate and discuss below: (i) binding energy, (ii) energies of the natural parity states ($s$- and $d$-wave resonances) in $^{16}$F ($^{15}$O+$p$ subsystem of $^{17}$Ne), (iii) interactions in $p$-waves in the $^{15}$O+$p$ channel (non-natural parity states of $^{16}$F), (iv) $s$-$d$ configuration mixing in $^{17}$Ne. These types of the dependence should be taken into account when we discuss our ability to reconstruct the low-energy radiative capture rates from nuclear experimental data.  Some of these types of the dependence have been discussed already  in Ref.\ \cite{Grigorenko:2006} but here we would like to provide a more systematic approach to the problem.

Note here, that the E1 non-energy weighted sum rule is connected with the structure of the ground state only. For calculations with fixed g.s.\ properties it remains the same, while the E1 strength can be strongly redistributed among different energy regions, crucially affecting the low-energy region important for radiative capture at astrophysical conditions. Thus for E1 strength function profiles, which are very similar in the typical experimentally observable range ($1-10$ MeV), essentially different low-energy behaviors are possible depending on dynamical peculiarities.

\begin{figure}[tb]
\begin{flushleft}
\includegraphics[width=0.254\textwidth]{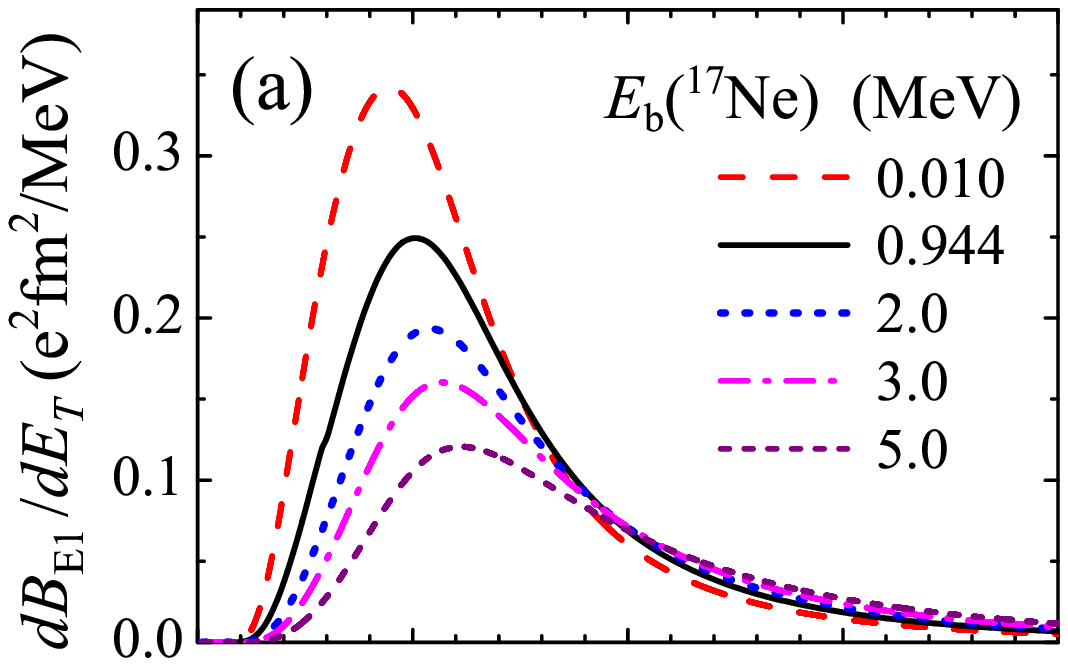}
\includegraphics[width=0.216\textwidth]{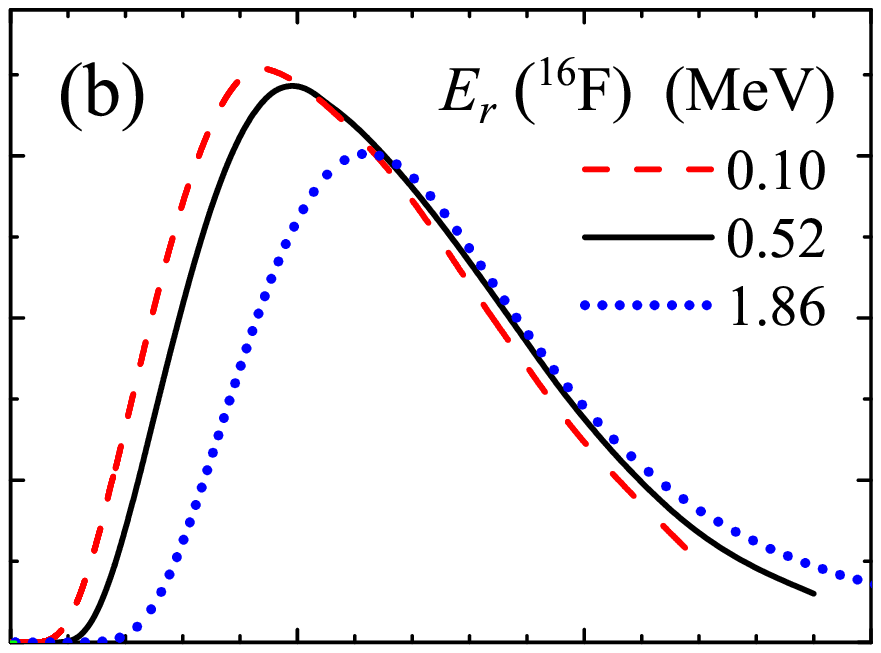}
\includegraphics[width=0.254\textwidth]{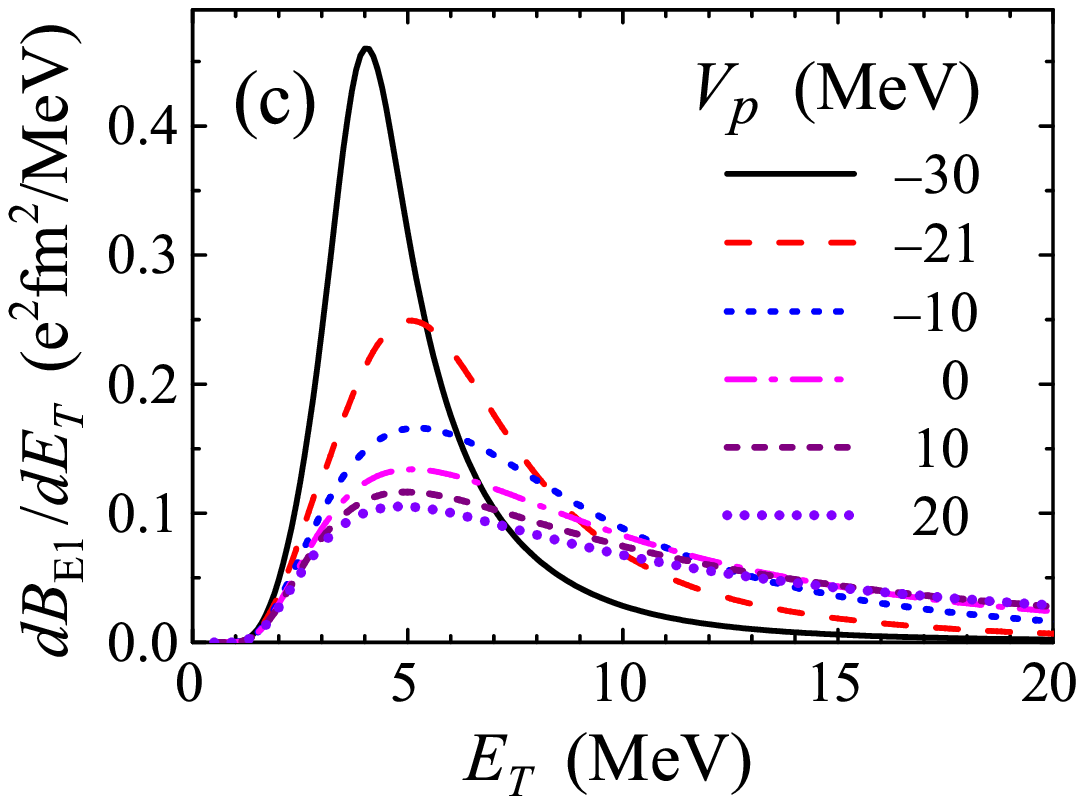}
\includegraphics[width=0.216\textwidth]{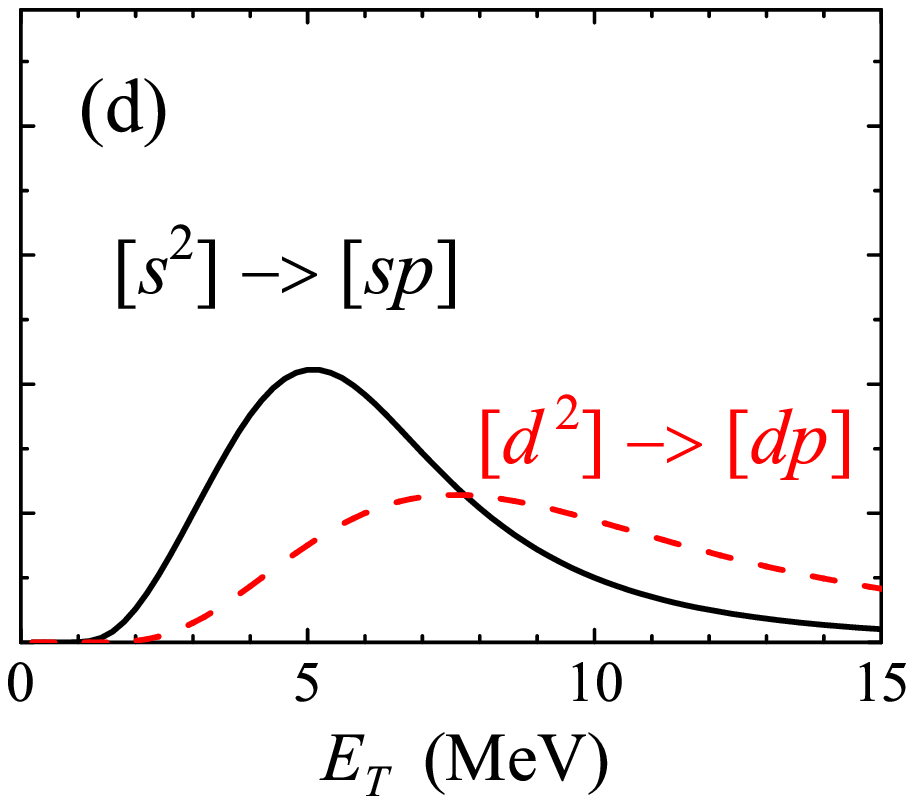}
\end{flushleft}
\caption{Dependence of E1 strength function in $^{17}$Ne g.s.\ on different parameters of the nuclear structure up to energy of several MeV. (a) Variation of the $^{17}$Ne g.s.\ binding energy. (b) Position of the low-lying $s$-wave resonance in $^{15}$O+$p$ channel (the g.s.\ energy of the $^{16}$F, which is a two-body subsystem of $^{17}$Ne). (c)  Interaction in the $p$-wave of the $^{15}$O+$p$ channel, see Eq.\ (\ref{eq:vp}) for parameter $V_p$. (d) The $[s^2]/[d^2]$ ratio in the $^{17}$Ne g.s.\ WF: $100\%$ of $[s^2]$ vs.\ $100\%$ of $[d^2]$.}
\label{fig:dep-all-lin}
\end{figure}


\subsection{Binding energy dependence}
\label{sec:calc-be-dep}


Figure \ref{fig:dep-all-lin} (a) shows the E1 strength function dependence on the binding energy of the three-body system. The binding energy variation for $^{17}$Ne is provided by artificial variation of the short-range three-body potential in the calculations, see Ref.\ \cite{Grigorenko:2003}. These E1 strength function variations are not of practical importance for the $^{17}$Ne case since the binding energy is well known to be $E_b = 0.933$ MeV \cite{Audi:2014}. However, the demonstrated trend is a nice illustration of the soft dipole mode of the E1 transition in $^{17}$Ne. Despite strong Coulomb interaction in $^{17}$Ne, which, in principle, is expected to suppress the SDM formation, the concentration of the E1 strength around $4-6$ MeV is stable in a broad range of $^{17}$Ne ``binding energies''. The E1 strength is strongly growing with $^{17}$Ne binding energy tending to zero and the peak position is moving towards lower energy.

Fig.\ \ref{fig:dep-all-lin} (a) shows that the dipole non-energy weighted (NEW) cluster sum rule is saturated for energies under $\sim 15$ MeV. There exists a well known connection between three-body cluster NEW sum rule and the root mean squared (rms) radius $r_3$ of the heavy core (with the mass number $A_3$ and the charge $Z_3$),
\begin{equation}
S_{\text{NEW}} = \int dE_T \, \frac{dB_{\text{E1}}}{dE_T} = \frac{3}{4\pi}  \, e^2 \, Z^2_{\text{eff}}  \,\langle r^2_3 \rangle \, ,
\label{eq:sum-rule-new}
\end{equation}
where $Z_{\text{eff}}=Z_3-A_3$ for $^{17}$Ne in $^{15}$O+$p$+$p$ model and $Z_{\text{eff}}=Z_3$ for $^{17}$N in the $^{15}$N+$n$+$n$ model. Thus, the sum rule in Fig.\ \ref{fig:dep-all-lin} (a) grows with the increase in the system size and consequently in the $\langle r^2_3 \rangle$ value.

\begin{figure}[tb]
\begin{flushleft}
\includegraphics[width=0.256\textwidth]{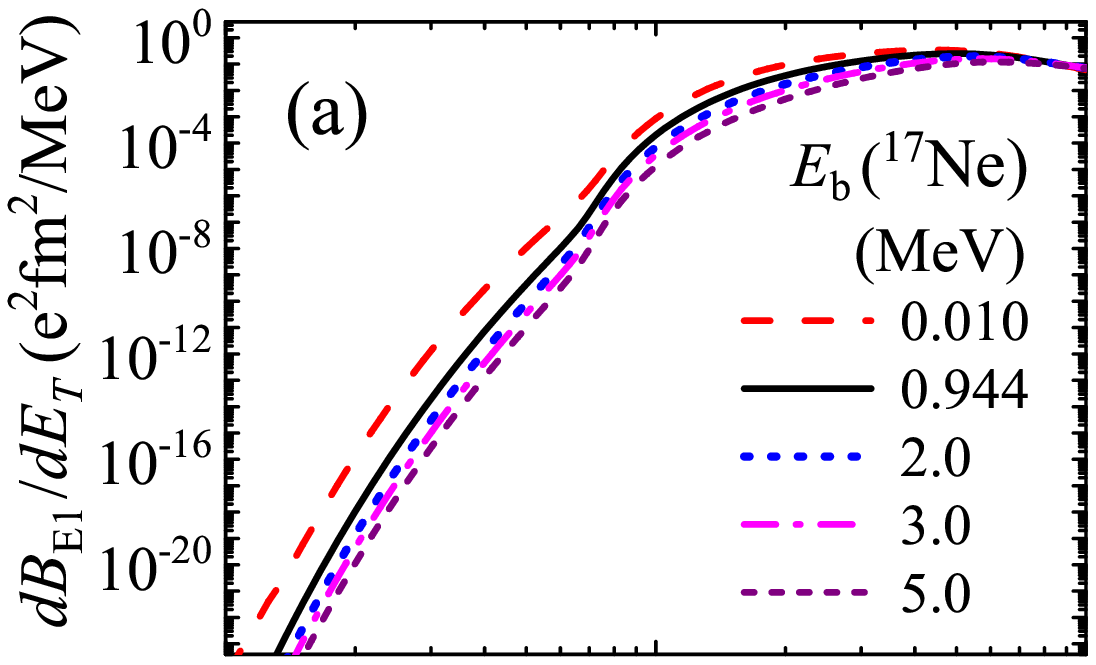}
\includegraphics[width=0.216\textwidth]{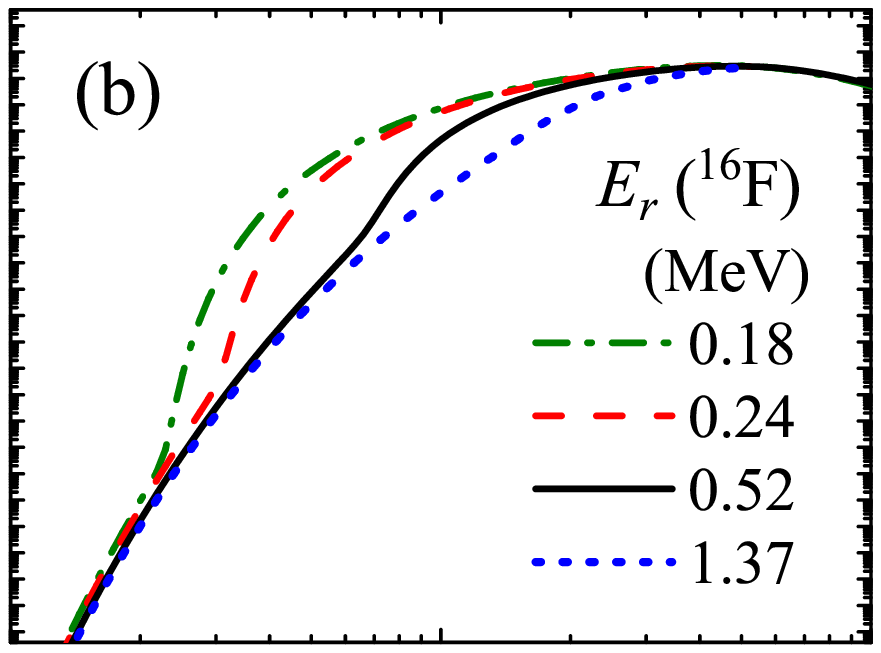}
\includegraphics[width=0.256\textwidth]{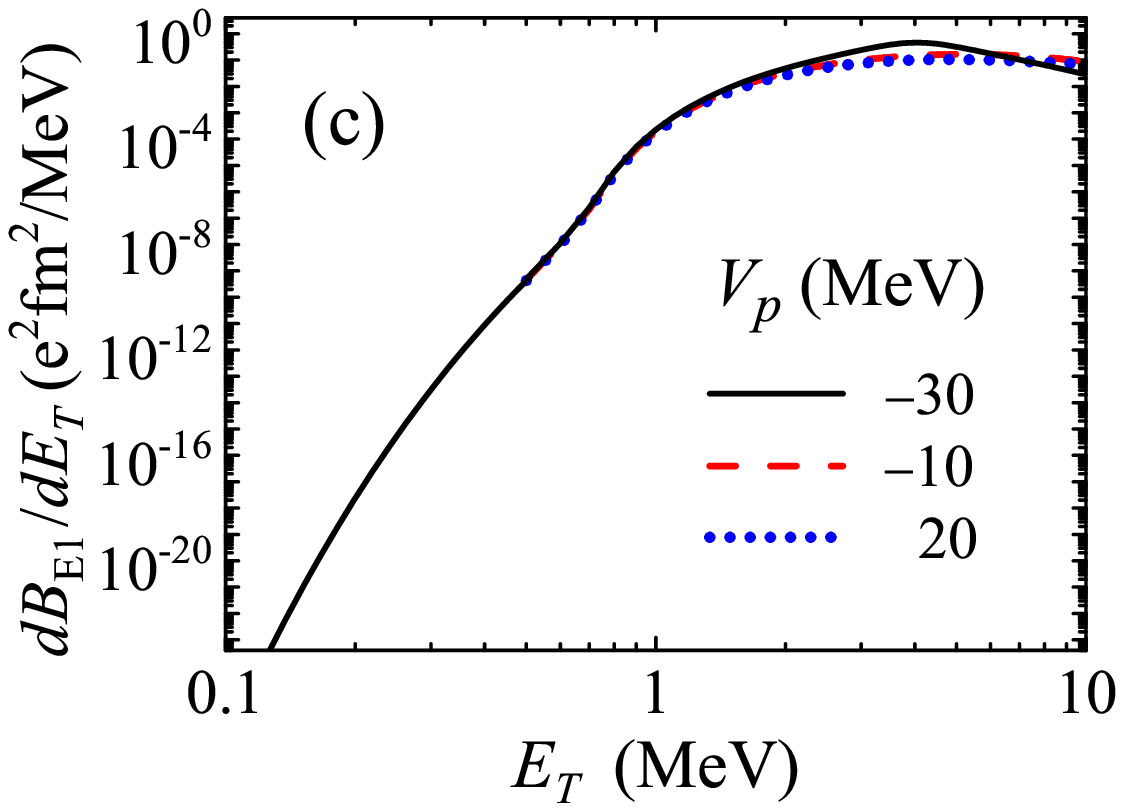}
\includegraphics[width=0.216\textwidth]{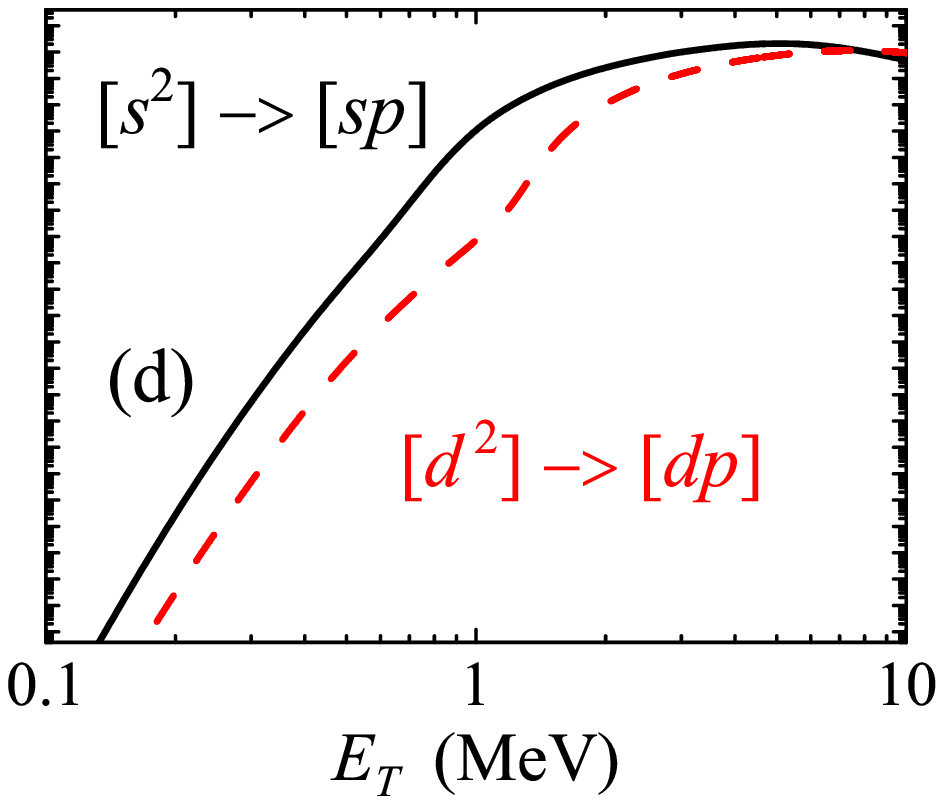}
\end{flushleft}
\caption{Sensitivity of E1 strength function in $^{17}$Ne g.s.\ to different aspects of nuclear structure at low energies. Panels are the same as in Fig.\ \ref{fig:dep-all-lin} but for extremely low energies (log scales).}
\label{fig:dep-all-log}
\end{figure}

The low-energy behavior of the E1 strength function is illustrated in Fig.\ \ref{fig:dep-all-log} (a). The low energy part of strength function varies by four orders of the magnitude when binding energy varies in the range $E_b = 0-5$ MeV. So, the g.s.\ binding energy is an essential parameter for calculations of quantities of astrophysical interest. In Figs.\ \ref{fig:dep-all-lin} (a) and \ref{fig:dep-all-log} (a) we see first illustration of the problem generally addressed in this work: how the shape variation in the experimentally measurable range above 1 MeV is transformed into variation of low-energy asymptotic of the strength function which defines the value of the astrophysical capture rate in the physically interesting range below $1-3$ GK.

\subsection{Dependence on the $s$-wave resonance energy in $^{16}$F  }
\label{sec:calc-s-dep}


The dependence, illustrated in Figs.\ \ref{fig:dep-all-lin} (b) and \ref{fig:dep-all-log} (b), is principal for determination of the low-energy E1 strength. This has already been discussed in Ref.\ \cite{Grigorenko:2006} but here we reproduce this discussion for completeness.

To make the point more straightforward we use a limited model with realistic $^{17}$Ne g.s.\ WF, but simplified $^{16}$F continuum with just one $s$-wave $0^-$ state in the $^{15}$O+$p$ channel (real resonance position $E_r= 0.545$ MeV).

In the linear scale of Fig.\ \ref{fig:dep-all-lin} (b) decrease in the $0^-$ state resonance energy leads to decrease in the maximum position of the E1 strength function. In Fig.\ \ref{fig:dep-all-log} (b) we can find more details about low-energy behavior. The first important thing is a kink in the low-energy region of the strength function located at about $E_T \approx 1.5 E_r$. This kink marks the transition from pure three-body direct capture mechanism at extreme low energies to ``quasi-binary'' capture mechanism at $E_T \gtrsim 1.5 E_r$. In the latter case the population of the intermediate $0^-$ resonance in the $^{15}$O+$p$ subsystem essentially enhances the E1 strength function compared to direct three-body capture regime.

One more thing should be noted here: the energy $E_r$ of the $s$-wave resonance in $^{16}$F affects not only the energies $E_T \gtrsim 1.5 E_r$ but also the extreme low-energies. The slope of the strength function is not affected for $E_T \lesssim 1.5 E_r$, but there is scaling factor, which is not that small. For example, for limiting cases of $E_r=0.1$ MeV and $E_r=1.86$ MeV in Fig.\ \ref{fig:dep-all-log} (b) the scaling factor $\sim 30$ can be found. This means that this aspect of the capture dynamics is affecting the astrophysical capture rate at the lowest temperatures.


\subsection{Dependence on the $p$-wave interaction in $^{16}$F}
\label{sec:calc-p-dep}

\begin{figure}
\begin{center}
\includegraphics[width=0.4\textwidth]{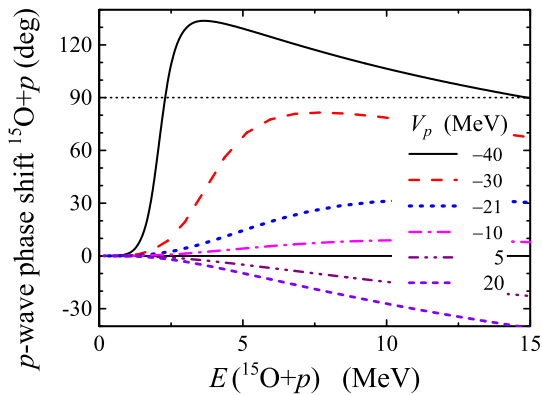}
\end{center}
\caption{The phase shifts in the $p$-wave of the $^{15}$O+$p$ channel for different $V_p$ parameters.}
\label{fig:dlt}
\end{figure}

Sensitivity of the E1 strength function to the $p$-wave interaction in the $^{15}$O+$p$ channel is illustrated in Figs.\ \ref{fig:dep-all-lin} (c) and \ref{fig:dep-all-log} (c). The following $p$-wave interaction of Woods-Saxon type with repulsive core was used, that of Ref.\ \cite{Grigorenko:2003}
\begin{equation}
V_p(r) =  \frac{V_p}{1+\exp[(r-2.94)/0.65} + \frac{200}{1+\exp[(r-0.89)/0.4} \,.
\label{eq:vp}
\end{equation}
Here we neglect a possible $ls$ component of the $p$-wave interaction. The originally employed value $V_p=-9$ MeV had no serious motivation, except the following: a small attraction, which does not contradict experimental data of $^{16}$F and $^{16}$N systems, where no low-lying positive parity (possibly $p$-wave) states are known. We vary the $V_p$ parameter in the range from $-30$ to 20 MeV, (from modest attraction to modest repulsion), see Fig.\ \ref{fig:dlt}. In reality the case $V_p=-30$ MeV is a borderline case with a phase shift not reaching $90^{\circ}$, but demonstrating resonance-like behavior at $E_r \sim 3.5$ MeV, which is seen as a quite sharp peak in the strength function Fig.\ \ref{fig:dep-all-lin} (c) at $E_T \sim 4$ MeV. For more attractive $p$-wave potentials a pronounced resonance is formed in the $^{15}$O+$p$ channel, see Fig.\ \ref{fig:dlt}.

Fig.\ \ref{fig:dep-all-lin} (c) and Table \ref{tab:sc-vp} show that the variation of the $p$-wave interaction leads to drastic modification of the E1 strength function in the energy range from 1 to 20 MeV. In contrast, if we zoom to the low-energy behavior in Fig.\ \ref{fig:dep-all-log} (c), no noticeable E1 strength function modification for $E_T<1$ is found. So, the astrophysical capture rate is not affected by this parameter for temperatures $T<2-5$ GK. However, the shape of the strength function in the energy range accessible for laboratory studies is essentially sensitive to this parameter.

\begin{table}[b]
\caption{Contributions of the energy region with $E_T$ less than the listed energy to the E1 Coulex cross section (in mb), calculated with different $p$-wave interactions in the $^{15}$O+$p$ channel (different $V_p$ parameters). Calculations with the realistic absorption function $F_{\text{abs}}$ and $^{17}$Ne g.s.\ WF with $\sim 50 \%$ of the $[s^2]$ configuration.}
\begin{ruledtabular}
\begin{tabular}[c]{ccccc}
$V_p$ (MeV)  & $<5$ MeV  &  $<7$ MeV &  $<10$ MeV & ``$\infty$''  \\
\hline
$-21$  & 148 & 243 & 297 & 322  \\
$-10$  & 109 & 179 & 236 & 277  \\
$0$    & 93  & 150 & 199 & 245  \\
\end{tabular}
\end{ruledtabular}
\label{tab:sc-vp}
\end{table}


\subsection{Dependence on the $s$-$d$ ratio}
\label{sec:calc-sd-dep}


The contributions of the dominant $[s^2]$ and $[d^2]$ components of the $^{17}$Ne g.s.\ WF to the E1 strength function are practically defined by incoherent $[s^2] \rightarrow [sp]$ and $[d^2] \rightarrow [dp]$ transitions.
It can be seen in Fig.\ \ref{fig:dep-all-lin} (d) that the $[s^2]$ contribution produces a peak in the strength function at $\sim 5$ MeV, while the $[d^2]$ --- at $\sim 8$ MeV. This information can, in principle, be used to extract the $[s^2]/[d^2]$ configuration mixing ratio. However, the previous section has shown that the effect of configuration mixing can be spoiled by the effect of the $p$-wave interaction [fixed $V_p=-21$ MeV is used for the calculations shown in Fig.\ \ref{fig:dep-all-lin} (d)]. For that reason these two types of dependence of the E1 strength function should be analyzed simultaneously in a broad energy range.

Sensitivity of the astrophysical capture rate to configuration mixing becomes clear from Fig.\ \ref{fig:dep-all-log} (d) (see also discussion in \cite{Grigorenko:2006}). In the energy range $E_T< 1$ MeV, the $[s^2] \rightarrow [sp]$ contribution to the strength function is larger than the $[d^2] \rightarrow [dp]$ contribution by (minimum) 3 orders of the magnitude. This means that the $[d^2] \rightarrow [dp]$ transition can become important only for weights of the $[s^2]$ configuration in the $^{17}$Ne g.s.\ WF less than $0.1 \%$, which is highly unrealistic situation.


\subsection{Qualitative discussion}


We have demonstrated above the dependence of the SDM strength function in $^{17}$Ne on four parameters. Two of these parameters (binding energy of $^{17}$Ne g.s.\ and resonance positions in $^{16}$F subsystem) are well fixed by experimental data. Two other parameters deserve special attention and are actually addressed in this work.

The parameter, connected with the structure of $^{17}$Ne (namely the $[s^2]/[d^2]$ configuration mixing ratio),  is well fixed in our calculations. These calculations are carefully tested against the experimental Coulomb displacement energy in the $^{17}$Ne and $^{17}$N \cite{Grigorenko:2003} and also electromagnetic characteristics ($B_{E2}$ values) \cite{Grigorenko:2005}. However, there exist alternative opinions about structure of $^{17}$Ne based on both theoretical and experimental considerations. There are theoretical works pointing out either $[s^2]$ \cite{Nakamura:1998,Timofeyuk:1996,Gupta:2002} or  $[d^2]$ \cite{Fortune:2001} domination in the structure of $^{17}$Ne. Note, the low-energy behavior of the E1 strength function is entirely defined by $[s^2]$ configuration and hence depends linearly on the $[s^2]/[d^2]$ configuration mixing ratio in $^{17}$Ne. As far as we can hardly expect that weight of the $[s^2]$ configuration is less than $5-10 \%$ under any assumptions about nuclear dynamics, the uncertainty of the astrophysical capture rate connected with the configuration mixing uncertainty is never larger than an order of magnitude.

The aspect, connected with intensity of the $p$-wave interactions (``non-natural'' parity states in the core+$p$ channel), has never been addressed in the literature so far. Here we note once again that variations of the $p$-wave interaction within reasonable limits are of negligible importance for E1 strength function in the energy range of astrophysical interest. However, it strongly affects the shape of the SDM above 1 MeV, see Fig. \ref{fig:dep-all-lin} (c) vs.\ Fig. \ref{fig:dep-all-log} (c). So, it becomes a factor of importance if we would like to ``extrapolate'' the strength function extracted from the Coulex data ($E_T \sim 1-10$ MeV) to astrophysical energies ($E_T<100$ keV). From theoretical point of view, such an uncertainty in the $^{17}$Ne case is connected with poor knowledge about positive parity states in the $^{15}$O+$p$ channel ($p$-wave continuum of the $^{16}$F system). It is clear that this uncertainty can be eliminated by experimential studies of the $^{15}$O+$p$ scattering.


\section{Comparison with experimental data}
\label{sec:comp-ex}


Recently, experimental data on three-body dissociation of a relativistic $^{17}$Ne beam on heavy (Pb) and light (C) targets have become available \cite{Marganiec:2016,Wamers:2018}. The data for the heavy target was interpreted in terms of Coulomb excitation. Publication \cite{Marganiec:2016} concentrated on the population of the low-energy resonant states of $^{17}$Ne, while the broad hump at $3-6$ MeV, which is interpreted as the SDM contribution in our present work, was discussed in Ref.\ \cite{Wamers:2018}.
The conclusion of \cite{Wamers:2018} is that the $3-6$ MeV hump is not consistent with any of the available theoretical E1 predictions. Below we use the experimental data obtained in \cite{Marganiec:2016,Wamers:2018}. We demonstrate below that actually the data are consistent with SDM prescription when all sensitivities to model parameters are taken into account. The data also allow to impose considerable limitations on the model parameters and, consequently, on the derived astrophysical rates.

\begin{figure}[tb]
\begin{flushright}
\includegraphics[width=0.250\textwidth]{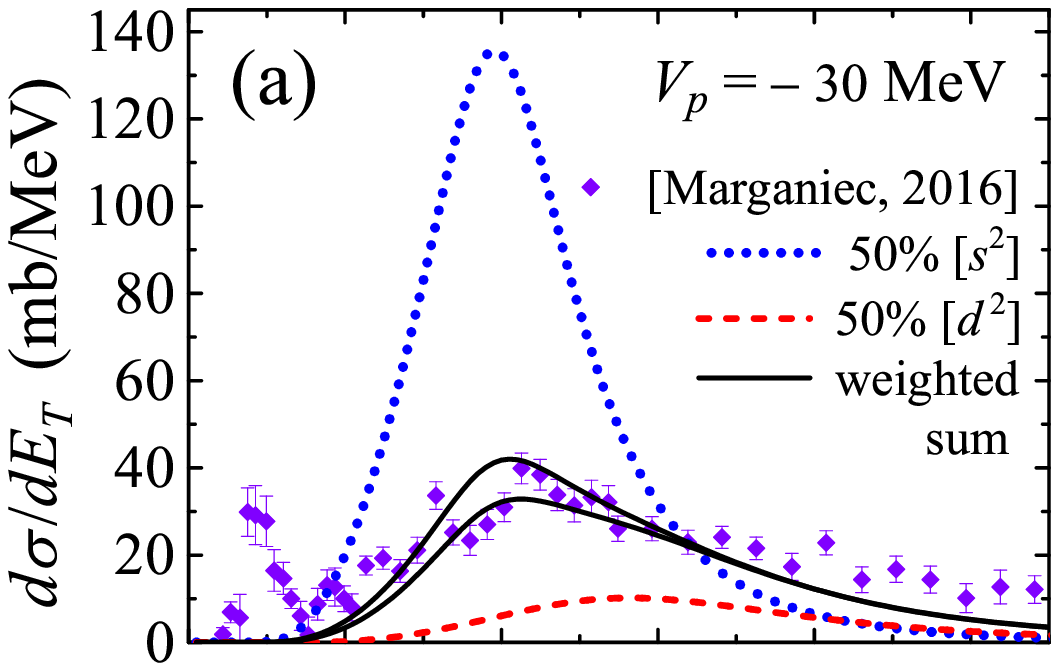}
\includegraphics[width=0.226\textwidth]{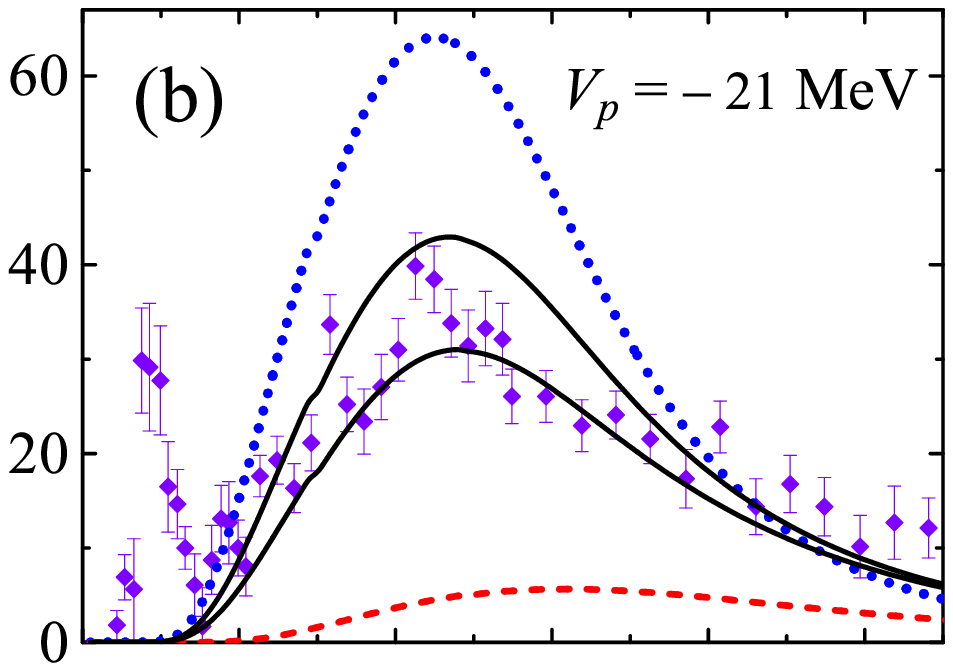}
\includegraphics[width=0.250\textwidth]{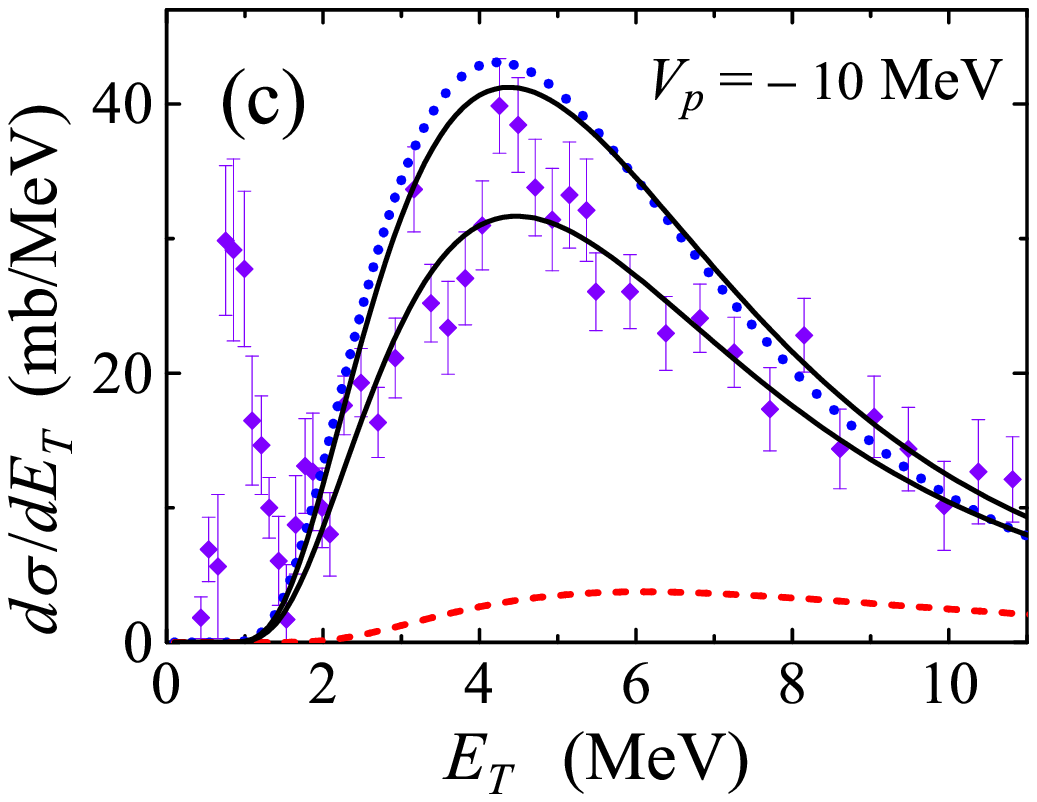}
\includegraphics[width=0.226\textwidth]{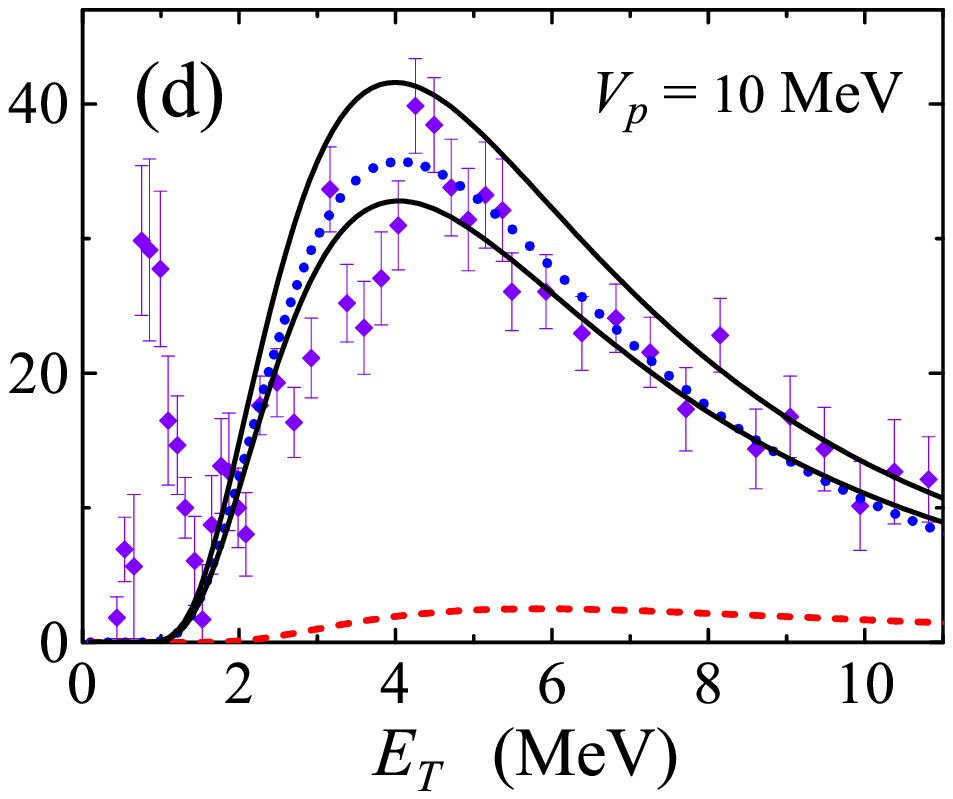}
\end{flushright}
\caption{Measured EMD cross sections (lead target, \cite{Marganiec:2016}) and those calculated with different potentials in the $p$-wave state in the $^{15}$O+$p$ channel. The blue dotted curves correspond to the contribution of the $50\%$  $[s^2]$ configuration of the $^{17}$Ne g.s., while the red dashed curves correspond to the contribution of the $50\%$ $[d^2]$ configuration. Two black solid curves are the upper and lower limit weighted sums of $[s^2]$ and $[d^2]$ fitting the experimental data. Note varying vertical scales.}
\label{fig:cs-sd-fit}
\end{figure}

We have shown above that there are two parameters which define the shape of the excitation function for energies above $\sim 1$ MeV and which are not fixed by experimental data. These are the $[s^2]/[d^2]$ ratio and the $p$-wave interaction in the $^{15}$O+$p$ channel. To fix both these parameters we use the following procedure. We start from the $^{17}$Ne g.s.\ WF from \cite{Grigorenko:2003,Grigorenko:2005} with $\sim 50 \%$ weight of the $[s^2]$ configuration. EMD cross sections are calculated for a broad range of $V_p$ parameters, see Fig.\ \ref{fig:dlt}. Then the cross section contributions stemming separately from the $[s^2]$ and the $[d^2]$ components of the  $^{17}$Ne g.s.\ WF are renormalized to fit the experimental data. In this way we derive the empirical $[s^2]/[d^2]$ configuration mixing for given $V_p$ value.

\begin{figure}[tb]
\begin{center}
\includegraphics[width=0.4\textwidth]{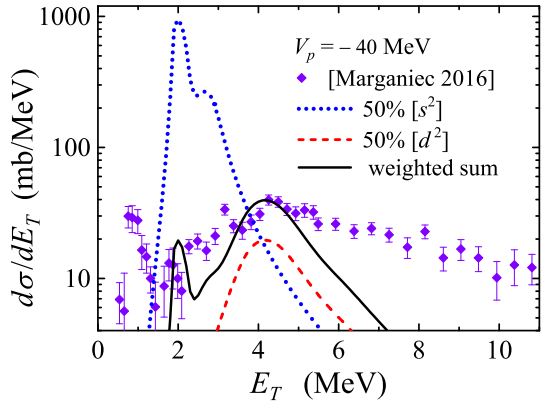}
\end{center}
\caption{Measured EMD cross sections and the one calculated with $V_p = -40$ MeV for potentials in the $p$-wave state in the $^{15}$O+$p$ channel for a lead target. The illustration of transition from SDM character of E1 strength function in Fig.\ \ref{fig:cs-sd-fit} to that generated on real resonant states. The weighted sum (black curve) supposes practically no contribution from the $[s^2]$ configuration to fit the data.}
\label{fig:cs-sd-res}
\end{figure}

This analysis is illustrated in Fig.\ \ref{fig:cs-sd-fit}. For the quite repulsive $p$-wave interaction of Fig.\ \ref{fig:cs-sd-fit} (d), the total calculated EMD cross section is small and therefore the large $[s^2] \rightarrow [sp]$ contribution is needed to fit the experimental data. For the attractive $p$-wave interaction of Fig.\ \ref{fig:cs-sd-fit} (a), shape of the $[s^2] \rightarrow [sp]$ contribution has a distinct peak. This marks the transition from SDM character of the E1 strength function to resonance behavior. In this case a minimal weight of the $[s^2]$ configuration in the $^{17}$Ne g.s.\ WF is sufficient to provide the intense E1 transition. For an even more attractive $p$-wave interaction a sharp resonance peaks arises in the strength function, see Fig.\ \ref{fig:cs-sd-res}. To match the data in this case only a negligible weight of the $[s^2]$ configuration in $^{17}$Ne g.s.\ WF ($\sim 2 \%$) should be assumed, which is highly unrealistic. Besides that the shape of the strength function is wrong.

\begin{figure}[tb]
\begin{center}

\includegraphics[width=0.37\textwidth]{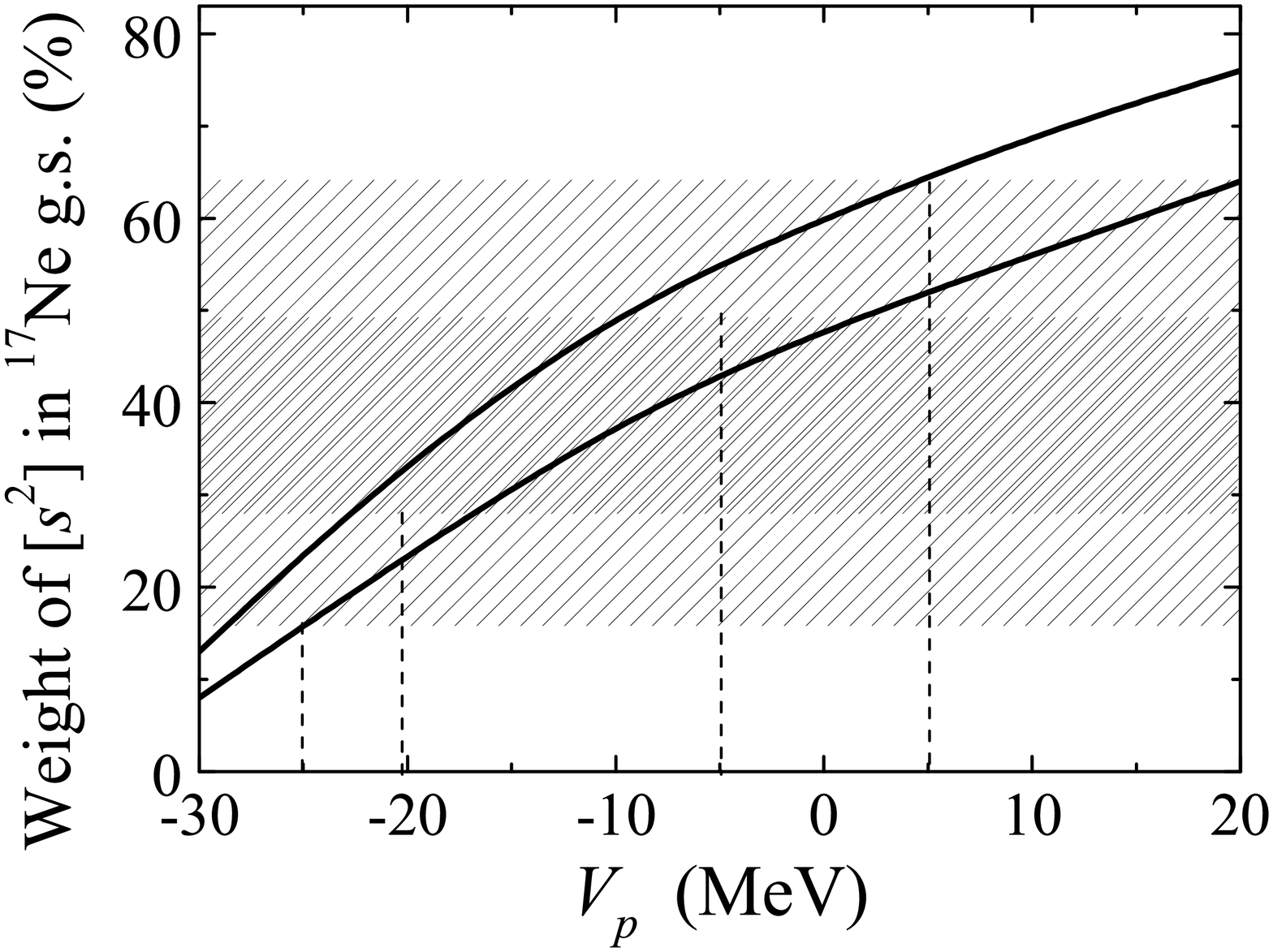}

\end{center}
\caption{The upper and lower limit weights of the $[s^2]$ configuration as obtained in the fits of Fig.\ \ref{fig:cs-sd-fit} with different $p$-wave interactions in the $^{15}$O+$p$ channel. The hatched ranges correspond to admissible $[s^2]$ weights.}
\label{fig:cs-sd-fit-sum}
\end{figure}

The results of this analysis are summarized in Fig.\ \ref{fig:cs-sd-fit-sum}. We have found the range of good fit to be from $V_p=-21$ MeV to $V_p=-5$ MeV (weak attraction), which corresponds to the $28-50$ $\%$ range of $[s^2]$ configuration in the $^{17}$Ne g.s.\ WF. Also we consider as tolerable fits obtained with $V_p=-30$ MeV to $V_p=5$ MeV (some attraction or some repulsion). This range provides more relaxed limitations of $15-65\,\%$ on the $[s^2]$ configuration. Thus the existing data on EMD of $^{17}$Ne generally support our earlier predicted (Refs.\ \cite{Grigorenko:2003,Grigorenko:2005}) structure of the $^{17}$Ne g.s.\ with $\sim 50 \%$ of the $[s^2]$ configuration.

\begin{figure}[tb]
\begin{center}
\includegraphics[width=0.247\textwidth]{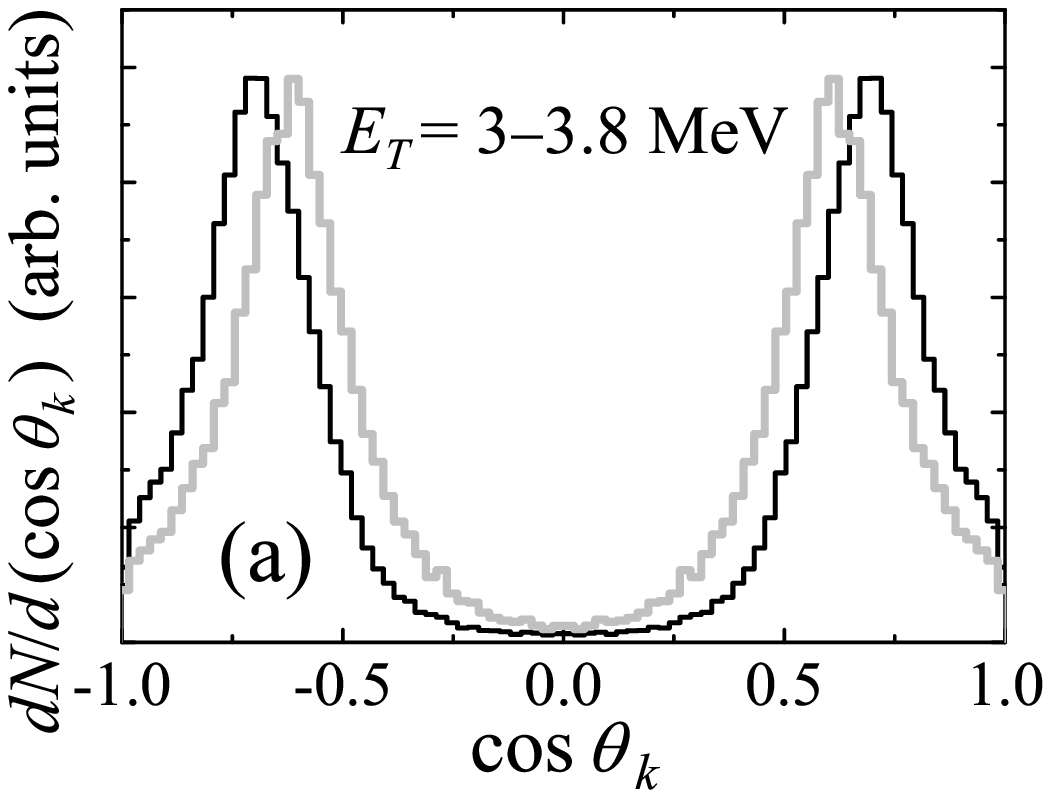}
\includegraphics[width=0.228\textwidth]{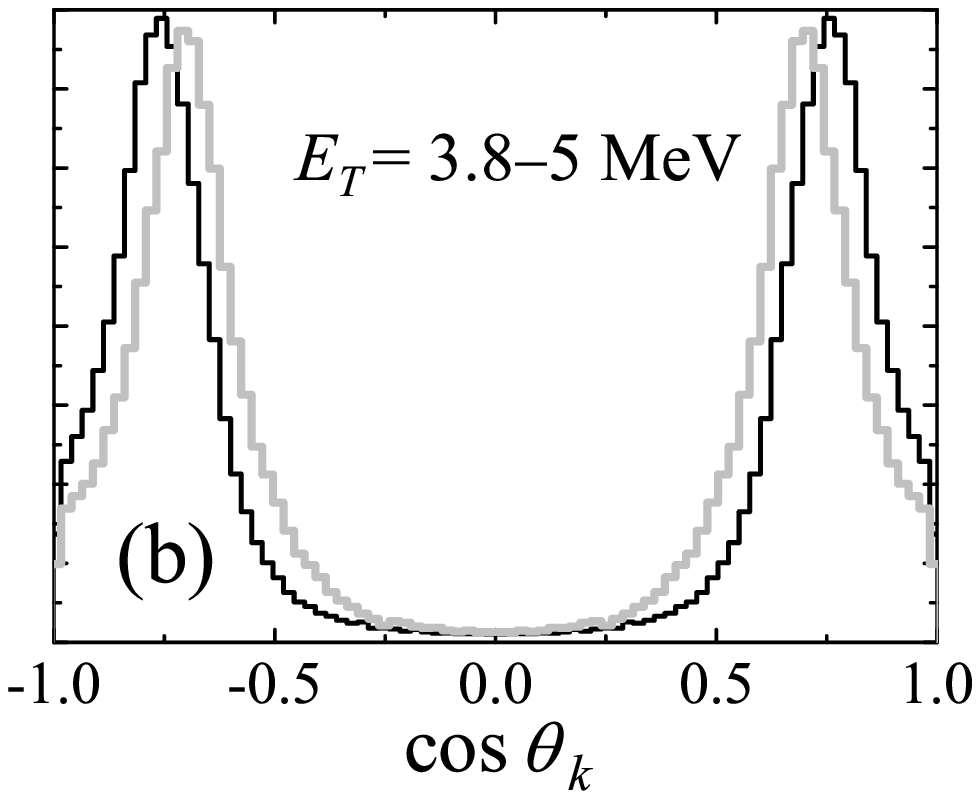}
\end{center}
\caption{Black and gray histograms show the Monte Carlo results for angular distribution in the ``Y'' Jacobi system for $[s^2] \rightarrow [sp]$ and $[d^2] \rightarrow [dp]$ transitions correspondingly. Case $V_p=-21$ MeV. Panels (a) and (b) correspond to left slope and peak region of the SDM strength function.}
\label{fig:ang-dis}
\end{figure}

Cross check of our conclusions on proposed description of the excitation spectrum of $^{17}$Ne as SDM can be obtained by studies of correlations between decay protons. The dissociation cross section \cite{Marganiec:2016} was obtained by invariant mass reconstruction, which means that the complete kinematical information including all possible three-body correlations are inherent in these data. Looking in Fig.\ \ref{fig:cs-sd-fit} (a)--(d) it is easy to see that the ratio of $[s^2]$ and $[d^2]$ components are quite different in provided fits. It should be understood that transitions $[s^2] \rightarrow [sp]$ and $[d^2] \rightarrow [dp]$ are dominating in these cases. So, excitation of $[s^2]$ component of $^{17}$Ne g.s.\ WF leads to population of the $s$-wave resonance states in the $^{15}$O+$p$ channel. These are the $0^-$ and $1^-$ states at 0.535 and 0.728 MeV, respectively, see Fig.\ \ref{fig:schemes}. Excitation of the $[d^2]$ component of the $^{17}$Ne g.s.\ WF leads to population of the $d$-wave resonance states: $2^-$ and $3^-$ at 0.959 and 1.256 MeV, respectively. Fig.\ \ref{fig:ang-dis} shows example how these population patterns are transformed into angular distributions of emitted protons. The distributions of Fig.\ \ref{fig:ang-dis} are obtained by the Monte-Carlo simulations taking into account the experimental bias in Ref.\ \cite{Marganiec:2016}. We can see that it seems realistic to disentangle contributions of the $[s^2] \rightarrow [sp]$ and the $[d^2] \rightarrow [dp]$ transitions at different three-body decay energies $E_T$.


\section{Comparison with theoretical calculations}

\begin{figure}[tb]
\begin{center}
\includegraphics[width=0.245\textwidth]{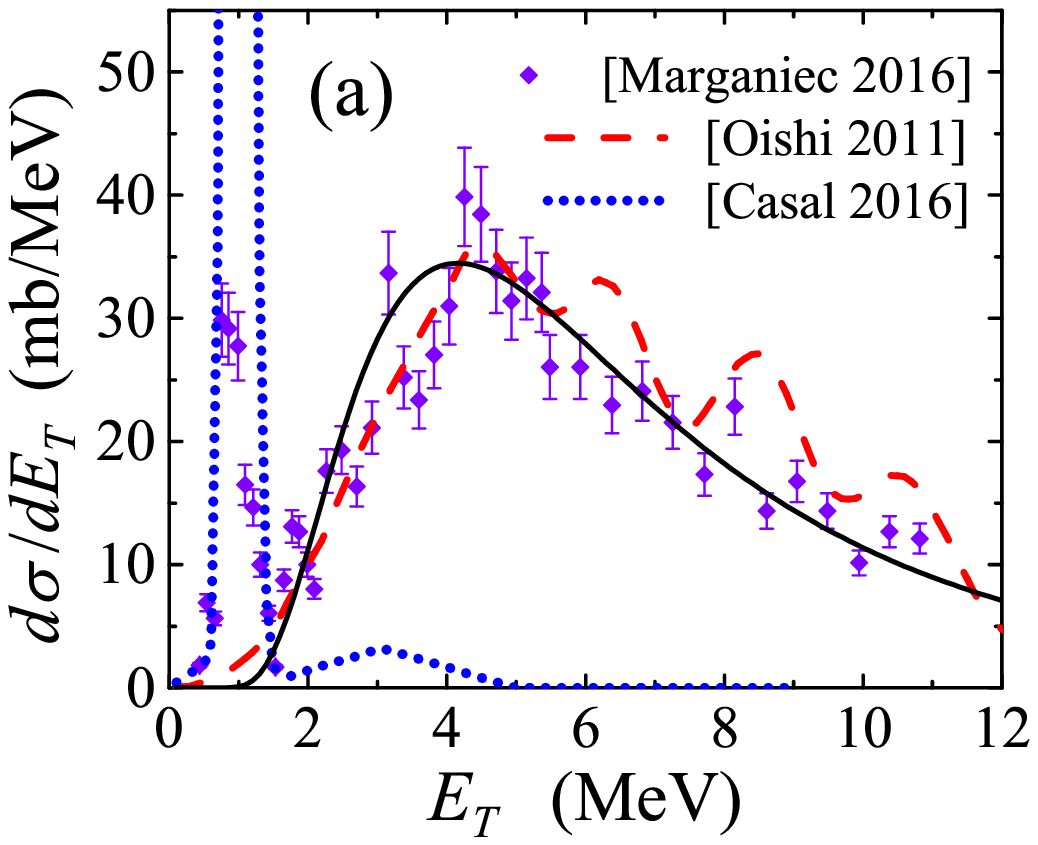}
\includegraphics[width=0.23\textwidth]{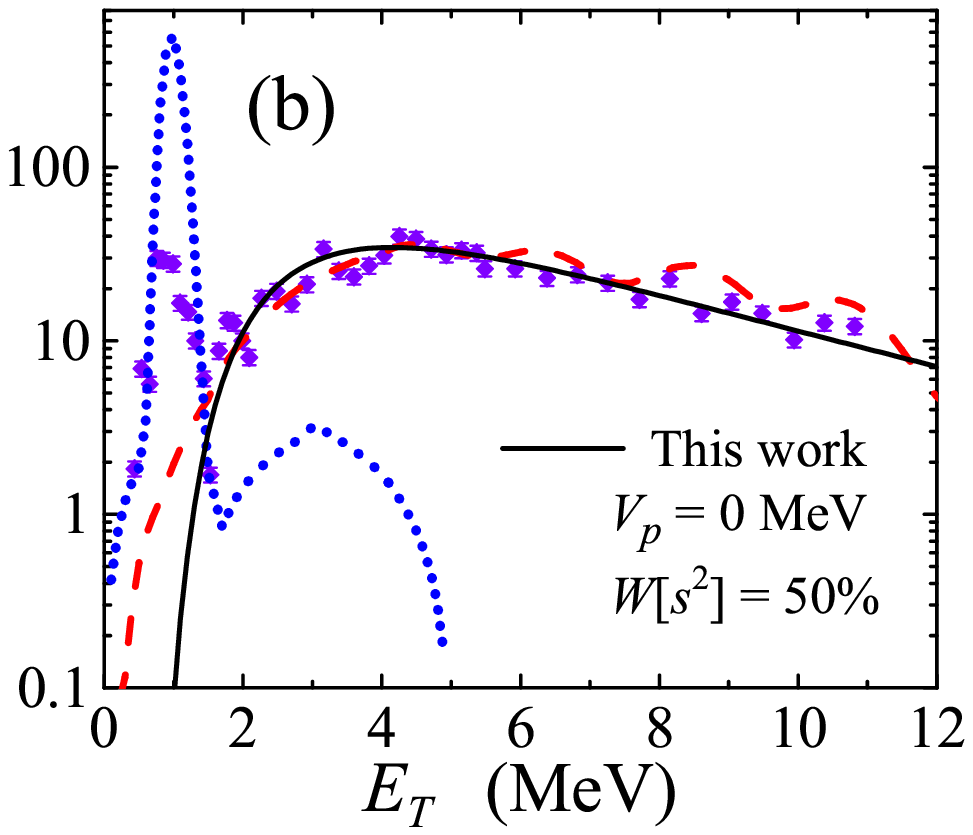}
\end{center}
\caption{Comparison of the theoretical E1 strength functions from Refs.\ \cite{Oishi:2011,Casal:2016} converted to Coulomb dissociation cross sections by Eq.\ (\ref{eq:bert-baur}) in linear (a) and logarithmic (b) scale. The diamonds shows the data (lead target) of Ref.\ \cite{Marganiec:2016}.}
\label{fig:th-comp}
\end{figure}

Since that time when our work \cite{Grigorenko:2006} appeared, the E1 strength function for $^{17}$Ne was twice discussed theoretically in the literature \cite{Oishi:2011,Casal:2016}. Figure \ref{fig:th-comp} compares the theoretical results after they are converted to the Coulomb dissociation cross section by Eq.\ (\ref{eq:bert-baur}).

The results of Oishi et al.\ \cite{Oishi:2011} are reasonably consistent with our results and with the experimental data \cite{Marganiec:2016}. The peak intensity is shifted somewhat to higher energies but not severely (say $\sim 1$ MeV) compared to our results. The cluster NEW sum rule for the energy range $E_T<12$ MeV which can be found from \cite{Oishi:2011} to be around $\sim 1.2$ $\text{e}^2\text{fm}^2$ compared to $\sim 1.57$ $\text{e}^2\text{fm}^2$ in our calculations, see Fig.\ \ref{fig:th-comp}. This could be evidence for higher $[d^2]$ configuration contents implied in the calculations of \cite{Oishi:2011}.

The wavy behavior in the predictions of \cite{Oishi:2011} is likely to be nonphysical since the curve is obtained by Gauss-smoothing of discrete-spectrum calculations. For the same reason the results of the paper \cite{Oishi:2011} are different from our results at energies below $\sim 1$ MeV. Is it clear that the E1 strength function obtained from the discrete spectrum with Gaussian smoothing do not possess correct low-energy asymptotic and thus it is not suited for calculation of astrophysical quantities at low temperatures.

In the case of the results of Ref.\ \cite{Casal:2016} we see, that the shape of the strength function is dramatically different from our predictions and from those of \cite{Oishi:2011}. We have to state that the strength function from Ref.\ \cite{Casal:2016} and all the conclusions based on it are erroneous. The main arguments here are the following:

\noindent (i) The EMD cross section predicted with the E1 strength function from \cite{Casal:2016} has little in common with experimentally observed picture. The peak at $E^*=1.76$ MeV comprises $\sim 180$ mb of integrated cross section. This by more than order of the magnitude exceeds the  experimental value from \cite{Marganiec:2016} $\sim 14.8(9)$ mb. It is highly improbable that such a massive contribution was missed in experiment.

\noindent (ii) The experimentally observed low-energy peak is ordinarily \cite{Chromik:2002,Marganiec:2016} interpreted as dominated by E2 excitation leading to the $5/2^-$ $^{17}$Ne state at $E^*=1.76$ MeV. The same conclusion is obtained in our present work where a $17.6$ mb E2 cross section for $5/2^-$ is predicted, see Table \ref{tab:crsec}, while the $E^*=1.76$ MeV peak with the strength function from Ref.\ \cite{Casal:2016} is attributed to the E1 cross section.

\noindent (iii) Some contribution to the 1.76 MeV peak from the E1 transition to the $1/2^+$ $^{17}$Ne state $E^*=1.908$ MeV is, in principle, possible. However, it is known from the studies of $^{17}$N, the mirror isobaric partner, that this state does not have a single-particle structure. Therefore the E1 strength for the $1/2^+$ state in $^{17}$N has extremely small strength $B_{\text{E1}} \sim 2.5 \times 10^{-6}$  $\text{e}^2\text{fm}^2$. The contribution of such a transition for Coulomb dissociation of $^{17}$Ne was estimated in Ref.\ \cite{Marganiec:2016} as $< 2.4$ mb with corresponding $B_{\text{E1}} < 7 \times 10^{-3}$  $\text{e}^2\text{fm}^2$.

\noindent (iv) We may guess about the origin of the low-energy peak in \cite{Casal:2016}. If we make the $p$-wave potential in the $^{15}$O+$p$ channel sufficiently attractive, a low-energy resonant state can be formed here. In such a case a low-lying single-particle positive parity state should arise in $^{17}$Ne built on the $[sp]$ configuration. The effect of such a state formation on the E1 EMD cross section is demonstrated in Fig.\ \ref{fig:cs-sd-res}. The calculation was performed with $V_p = -40$ MeV and corresponding resonance energy is $E_r \sim 2.3$ MeV, see Fig.\ \ref{fig:dlt}. Figure \ref{fig:cs-sd-res} illustrates the transition from SDM E1 strength function (see Fig.\ \ref{fig:cs-sd-fit}) to resonant strength function with corresponding abnormal E1 EMD cross sections. From a theoretical point of view the prerequisite of such a transition is existence of a resonant $p$-wave state in the $^{15}$O+$p$ channel with the resonance energy $E_r \lesssim 3$ MeV. Existence of such states contradicts known experimental spectra of $^{16}$F and its isobaric mirror $^{16}$N system.


\section{Astrophysical radiative capture rate}

\begin{figure}[tb]
\begin{center}
\includegraphics[width=0.45\textwidth]{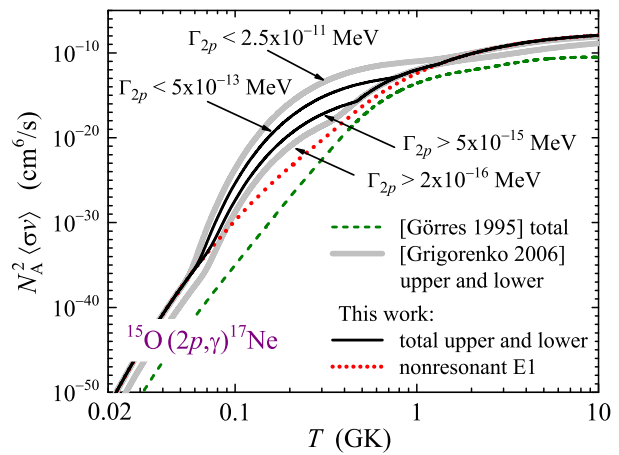}
\end{center}
\caption{Astrophysical radiative capture rate for the $^{15}$O+$p$+$p\rightarrow ^{17}$Ne+$\gamma$ reaction. The results of this work are compared with of our previous work \cite{Grigorenko:2006} and Ref.\ \cite{Gorres:1995}.}
\label{fig:rate}
\end{figure}

The \emph{nonresonant} astrophysical radiative capture rate for the $^{15}$O+$p$+$p\rightarrow ^{17}$Ne+$\gamma$ reaction was calculated using Eq.\ \ref{eq:nonres-rate} and E1 strength functions obtained for admissible values of the $[s^2]/[d^2]$ configurations mixing from $15 \%$ to $65 \%$, see Fig.\ \ref{fig:cs-sd-fit-sum}. The obtained rate is shown by a dotted curve in
Fig.\ \ref{fig:rate}. There is only one curve in the figure, as due to the scale of the vertical axis the difference between the cases of upper and lower limits of the $[s^2]/[d^2]$  mixing is comparable to the thickness of the curve. This is considerable improvement compared to our previous results of Ref.\ \cite{Grigorenko:2006}: There a comparatively broad uncertainty band for nonresonant capture rate was connected to assumption about principal  possibility for very low weights of $[s^2]$ configurations ($<5\%$) in the structure of $^{17}$Ne g.s.\ WF. Such a possibility is ruled out in the present work.

The \emph{resonant} astrophysical radiative capture rate for the $^{15}$O+$p$+$p\rightarrow ^{17}$Ne+$\gamma$ reaction was calculated according to the following equation, see Ref.\ \cite{Grigorenko:2005a} :
\begin{eqnarray}
\left\langle \sigma _{2p,\gamma }v\right\rangle & = & \left( \frac{A_1+A_2+A_3}{A_1 A_2 A_3}\right)^{3/2} \left( \frac{2\pi }{mkT}\right) ^{3}\sum_n \frac{2J_f(n)+1}{2(2J_i+1)} \nonumber
\\
 & \times & \, \exp \left[ -\frac{E_T(n)}{kT}\right] \frac{\Gamma_{2p}(n)\Gamma_{\gamma}(n)}{\Gamma(n)}  \,, \qquad
\label{eq:res-rate}
\end{eqnarray}
where $E_T(n)$, $J_f(n)$, $\Gamma(n)$, $\Gamma_{2p}(n)$, and $\Gamma_{\gamma}(n)$  are, respectively, two-proton decay energy, total spin, total two-proton and $\gamma$ widths of the $n$-th resonance in the spectrum of $^{17}$Ne. Parameters of the considered resonances can be found in Table II of Ref.\ \cite{Grigorenko:2005a}, except for the case of the first excited $3/2^-$ state of $^{17}$Ne. The two-proton width of this state defines the total astrophysical radiative capture rate in the temperature range from $\sim 0.07$ to $\sim 0.7$ GK. There are two updates of this value since the publication \cite{Grigorenko:2005a}. The first update concerns the theoretically predicted $2p$ width of this state which was found to be in the range $(5-8)\times 10^{-15}$ MeV in Ref.\ \cite{Grigorenko:2007}. The value $5\times 10^{-15}$ MeV is accepted as a \emph{lower} limit for $\Gamma_{2p}(3/2^-)$ in this work instead of $2\times 10^{-16}$ MeV accepted in Ref.\ \cite{Grigorenko:2005a}. The second update concerns the recent experimentally derived \emph{upper} limit for $2p$ width of this state \cite{Sharov:2017} which was established to be $\sim 50$ times lower than the previous upper limit from Ref.\ \cite{Chromik:2002}. The value $5\times 10^{-13}$ MeV is accepted as an \emph{upper} limit for $\Gamma_{2p}(3/2^-)$ in the present work instead of $2.5\times 10^{-11}$ MeV accepted in \cite{Grigorenko:2005a}. These changes induce a shrinking of the uncertainty range for the rate compared to uncertainty range in Ref.\  \cite{Grigorenko:2005a}, for the temperatures from $\sim 0.07$ to $\sim 0.7$ GK, see Fig.\ \ref{fig:rate}.

\begin{figure}[tb]
\begin{center}
\includegraphics[width=0.45\textwidth]{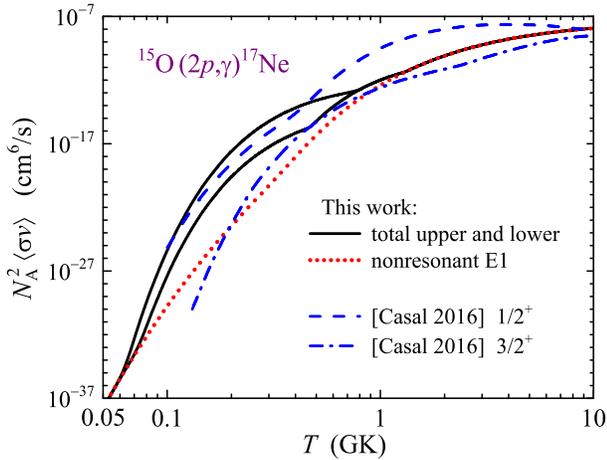}
\end{center}
\caption{Astrophysical radiative capture rate for the $^{15}$O+$p$+$p\rightarrow ^{17}$Ne+$\gamma$ reaction. The results of this work are compared with Ref.\ \cite{Casal:2016}.}
\label{fig:rate-comp}
\end{figure}

The calculated astrophysical radiative capture rates provided in Ref.\ \cite{Casal:2016} for the $1/2^+$ state and temperatures $T<5$ GK are $2-3$ orders of the magnitude larger than our results from Ref.\ \cite{Grigorenko:2006} and those obtained in the present work, see Fig.\ \ref{fig:rate-comp}. This excess is evidently connected with the contribution of the $\sim 1$ MeV peak in the E1 strength function predicted in \cite{Casal:2016}. Such a peak is not tolerated by the experimental data, as discussed above, which disqualifies the results of \cite{Casal:2016} for the astrophysical radiative capture calculations. We should note that the rates provided in \cite{Casal:2016} do not follow any trend of our rates neither for the $3/2^+$ component, nor for the $1/2^+$ component in any temperature range.

\section{Conclusions}


In the $^{17}$Ne nucleus possible existence of the $2p$ halo structure of the g.s.\ and the related soft dipole mode in the continuum are issues of serious current interest. In this work, we have discussed the major qualitative properties of the soft dipole mode in $^{17}$Ne and its relevance for determination of low-energy cross sections used for derivation of the astrophysical radiative capture rates. We demonstrate that ``extrapolation'' of Coulomb dissociation information for three-body systems to extremely low-energies, is a more complicated task than for two-body systems. The parameters which are needed to be fixed to accomplish this task are determined.

It should be understood that the general features of the soft dipole mode in three-body (core+$2N$) systems are well illustrated by the $^{17}$Ne example. The consideration of such a process in typical even $sd$-shell system can be done absolutely stereotypically and extension to $pf$-shell systems is easily done by analogy. The problems of ``extrapolation'' from the energy range, where the E1 strength function can be obtained from the Coulex experiment (few MeV), to low energies (under and around 1 MeV), important for the capture rate calculations, should be more or less the same and this work shows how they can be resolved.

We have shown that recently available Coulomb dissociation data for the $^{17}$Ne \cite{Marganiec:2016,Wamers:2018} are well described assuming a soft dipole mode for $^{17}$Ne continuum. These data allow to constrain the $[s^2]/[d^2]$ ratio in the structure of $^{17}$Ne to $27-50 \%$ of the $[s^2]$ configuration ($15-65 \%$ for a more relaxed fit). We demonstrate in this work that these limits can be further improved by studies of the core+$p$ correlations in the SDM spectrum of the $^{17}$Ne and the positive parity states in the spectrum of $^{16}$F.

We confirmed the non-resonant radiative capture rate from Ref.\ \cite{Grigorenko:2006} and considerably improved the uncertainty range for them, based on the recent experimental and theoretical results. We have demonstrated that the recent predictions of Ref.\ \cite{Casal:2016} are erroneous since they strongly contradict the experimental Coulomb dissociation data of Ref.\ \cite{Marganiec:2016,Wamers:2018} and the conceptual understanding of structure and excitations of the $^{17}$Ne and $^{17}$N isobaric partners.


\section{Acknowledgements}


Yu.L.P.\ and L.V.G.\ were partly supported by the Russian Science Foundation (grant No.\ 17-12-01367). I.A.E.\ was supported by the Helmholtz Association under grant agreement IK-RU-002.


\bibliographystyle{apsrev4-1}
\bibliography{d:/latex/all}


\end{document}